\documentclass[%
amsmath,amssymb,onecolumn,%
superscriptaddress,11pt,floatfix
]{revtex4-1}

\usepackage{graphicx}

\begin{document}







\title{A Comparison and Unification of Ellipsoidal Statistical and Shakhov BGK Models}

\author{Songze Chen}
\affiliation{
Hong Kong University of Science and technology \\
Clear Water Bay, Kowloon, Hong Kong, China
}%
\author{Kun Xu}%
\email{makxu@ust.hk}
\affiliation{
Hong Kong University of Science and technology \\
Clear Water Bay, Kowloon, Hong Kong, China
}%
\affiliation{
LTCS and CAPT, \\ Department of Mechanics and Aerospace Engineering, \\
College of Engineering, Peking University, Beijing 100871, China
}%

\author{Qingdong Cai}
\affiliation{
LTCS and CAPT, \\ Department of Mechanics and Aerospace Engineering, \\
College of Engineering, Peking University, Beijing 100871, China
}%

\date{\today}

\begin{abstract}
The Ellipsoidal Statistical model (ES-model) and the Shakhov model (S-model) are constructed
for the correction of Prandtl number of the original BGK model through the modification of stress and
heat flux. Even though in the continuum flow regime, both models can give the same Navier-Stokes equations with
correct Prandtl number, their modification of the collision term may have different dynamic effect in the non-equilibrium
transition flow regimes.
With the introduction of one free parameter, a generalized kinetic model with the combination of the ES-model and S-model can be
developed, and this new model can get the correct Navier-Stokes equations in the continuum flow regime as well, but with
abundant dynamic effect through the adjustment of the new degree of freedom.
In order to validate the generalized model, a numerical method based on the unified gas kinetic scheme (UGKS) has been developed for the
new model.
The physical performance of the new model with the variation  of the free parameter has been tested, where the ES-model and S-model become
the limiting cases.
In transition flow regime, many physical problems, i.e., the shock structure and micro-flows, have been studied using the generalized model.
With a careful choice of the free parameter, good results can be achieved for most test cases.
The overall conclusion is that the S-model
predicts more accurate numerical solutions in most tough test cases presented in this paper than the ES-model,
while ES-model performs better in the cases when the flow is mostly driven by heat, such as a channel flow with large boundary temperature variations
at high Knudsen number.
The numerical study demonstrates the necessity of developing such a generalized model.
With the inclusion of one more freedom, in the transition regime the new kinetic model may provide more accurate solution than the ES and Shakhov models.

\end{abstract}

\maketitle


\section{Introduction}
\label{sec:introduction}

The monatomic rarefied gas behavior can be described by the Boltzmann equation.
However, the collision term of the Boltzmann equation is a multiple integral term
which is very complicated for analysis and numerical computation.
The kinetic model is a simplification of the Boltzmann equation. The
simplest kinetic model is the BGK model \cite{BGK1954} in which the collision term is replaced
by a relaxation term. This relaxation term in the BGK model mimics the main relaxation process from nonequilibrium state
towards to a local equilibrium one with a Maxwellian distribution function.
The local equilibrium state is determined by the local conservative flow variables, namely, the density, the momentum and the energy.
The BGK model becomes an important kinetic model for analysis and numerical simulation of nonequilibrium flows.
However, the Chapman-Enskog expansion of the BGK model gives the Navier-Stokes equations with a unit Prandtl number, which
is different from the physical reality in the continuum flow regime. For a monatomic gas, the accepted Prandtl number is about $ 2/3$ in a
wide range of flow conditions.

In order to fix the Prandtl number, many kinetic models have been proposed in the past decades.
The main idea is to modify the relaxation states.
For example, the Ellipsoidal Statistical BGK model \cite{Holway1966} employs a Gaussian distribution as the
relaxation equilibrium state instead of the Maxwellian.
This model is not very popular until the proof of the entropy condition by Andries \emph{et. al.} \cite{Andries2001}.
In the ES-model, besides the conservative flow variables, the local stress tensor also involves in the post-collision state.
By changing the free parameter in the ES-model, it can present an arbitrary Prandtl number.
Moreover, the nonnegative property of the Gaussian distribution becomes a favorable physical property.

Another very popular kinetic model is the Shakhov model \cite{Shakhov68}. Unlike the ES-model, it
adjusts the heat flux in the relaxation term, but keeps the stress tensor the same as the original BGK  one.
The Hermit polynomial is adopted to modify the heat flux. So in terms of low order moments, the S-model
keeps the same as the BGK model.
The S-model also presents a correct Prandtl number. But, it allows negative value of distribution function,
and its H-theorem was only proved in near equilibrium condition \cite{Shakhov68}.

In 1990, Liu proposed a new kinetic model by considering the gain term and
lost term  of the Boltzmann equation separately \cite{Liu1990}. He used the Chapman-Enskog
distribution directly to evaluate the relaxation term. The modification of the collision term involves
the space derivatives. Liu model changes both the heat flux and stress tensor of the relaxation process,
and provides a correct Prandtl number in the continuum flow regime. Due to its
relatively complicated formulation, this model has not been widely used.

Although all above models provide
correct Prandtl number in the continuum flow regime, their properties are very different
in the transition regime \cite{Kudryavtsev2008,Zheng2005,Andries2002,Garzo1994,Graur2009}.
Garz\'{o} reported a singular behavior of Liu model and attributed it to the negative
distribution function \cite{Garzo1994}.  Graur studied the heat transfer problem,
and found that the ES-model provides better results than the S-model through the comparison with
the results from the Boltzmann equation.
The ES-model keeps
the distribution function positive, while the S-model and Liu model always allow
un-physical negative distribution function. It seems that the nonnegative properties
of the ES-model are important and promising. Moreover, the ES-model satisfies the H theorem, while the
H-theorem of the S-model is only proved in the near local equilibrium state \cite{Zheng2005}.
However, some other studies did not tell the same story. Mieussens \cite{Mieussens2004}
and Kudryavtsev \emph{et al.} \cite{Kudryavtsev2008} both reported the early rising of
temperature profile in the shock structure solution by the ES-model.

In fact, the physical performance of these models
has not yet been evaluated extensively in the transition regime.
The properties, such as the H-theorem, nonnegative distribution,
and conservation etc., cannot cover a complete picture of dynamics of the particle collision term and the evolution of the
distribution function.
Since the original motivation for the development of the kinetic models is to fix the Prandtl number which is well
defined in the continuum flow regime, in transition regime it is expected  that significant differences in their performance
would appear in different physical problems.
Furthermore, the practical requirement cares more about the
macroscopic quantities, such as the moments of a distribution function. The H-theorem and
the nonnegative distribution function cannot guarantee a correct dynamic  evolution of macroscopic quantities.
So, it is necessary to inspect the practical performance of different kinetic models
through the numerical simulations in the transition regime.
In order to cover a whole spectrum of dynamic performance of kinetic models,
we are going to introduce a generalized kinetic model which combines the ES-model and S-model.
With the combination of these two models, besides the correct capturing of Prandtl number in the new model, we have one more
free parameter to be adjusted. With the variation of this parameter, a continuum dynamic performance from ES-model to S-model, and beyond, can be
identified.

In the past years, a unified gas kinetic scheme (UGKS)  \cite{ugks1,ugks1_1,ugks2}
has been well developed. The BGK model and
the S-model have been employed in the UGKS. In this paper we will use  the
UGKS framework to construct numerical scheme for the generalized kinetic model.
The numerical scheme will be used to exam physical performance of different kinetic models with the variation of the parameter, where
both the ES-model and S-model become limiting cases.
A continuous dynamic transition between these two models can be obtained.
Through investigations, the  performances of different kinetic models in the transition regime are
presented in details.

This paper is organized as following. Section 2
presents the UGKS for the ES-model and other kinetic models. Section 3 proposes a generalized kinetic model.
Section 4 gives the simulation results of the new model in the shock structure and microflow computations.
The parameter dependent dynamic effect will be discussed in different test cases.
Section 5 presents the analysis and insight of the new model.
The last section is the conclusion.

\section{Unified gas kinetic scheme for kinetic models}
\label{sec:scheme}
The unified gas kinetic scheme is a direct modeling method to simulate flows in the whole Knudsen number regimes.
 It is a finite volume conservation law for the evolution of gas distribution function.
Besides the evolution of conservative flow variables,
such as density, momentum and energy, the time  evolution
of gas distribution function at discrete particle velocity is solved as well
in order to capture the non-equilibrium molecular transport.
Therefore, how to evaluate the fluxes of a gas distribution function across a cell interface is a central ingredient in UGKS.
The kinetic model is always employed in UGKS to provide the evolution dynamics of the distribution function, but the UGKS
is not targeting to solely solve the kinetic model itself, because the physical modeling scale of the kinetic model can be
different from the numerical cell size scale. The UGKS is a direct physical modeling of flow motion in the scale of the discretized space and
the integral solution used from the kinetic model covers the flow evolution from kinetic to the hydrodynamics scales. The specific flux
used at the cell interface depends on the ratio of time step to the local particle collision time.

In this section, a brief review of the
UGKS is presented.
Since the Shakhov model has been implemented in UGKS, this section will introduce the UGKS with a general kinetic model.
Generally, a kinetic model takes the following formulation,
\begin{equation}
    \frac{\partial f}{\partial t}+\mathbf{u}\cdot\frac{\partial f}{\partial \mathbf{x}} =
    \frac{g^{+}-f}{\tau}.
    \label{eq:kineticModel}
\end{equation}
The $f$ represents the velocity distribution function, and the $g^+$ is the post collision term.
The $\mathbf{x}$ and $\mathbf{u}$ represent the physical space variables and the velocity space variables
respectively. Here $\tau$ is relaxation time.

The macroscopic quantities, such as, the mass $\rho$, momentum $\rho \mathbf{U}\ (\rho U_{i})$, energy $\rho E$,
stress tensor $\mathbf{P}\ (p_{ij})$ and heat flux $\mathbf{q}\ (q_i)$,
can be derived from the distribution function $f$,
\begin{eqnarray}
  W = \left(\begin{array}{l} \rho \\ \rho \mathbf{U} \\ \rho E \end{array}\right)
    &=& \int \mathbf{\psi} f d \mathbf{u}, \nonumber\\
  p_{ij} &=& \int (u_i-U_i)(u_j-U_j)fd\mathbf{u}, \label{eq:macroMicroRelation} \\
  q_{i} &=& \int \frac{1}{2}(u_i-U_i)(\mathbf{u}-\mathbf{U})^2fd\mathbf{u} \nonumber,
\end{eqnarray}
where $\mathbf{\psi}$ is defined as following,
\begin{equation}
  \mathbf{\psi} = (1,\mathbf{u},\frac{1}{2}\mathbf{u}^2)^T,
\end{equation}
and $d\mathbf{u}$ is the volume element in the velocity space.
Since mass, momentum, and energy are conserved during particle collisions,
$f$ and $g^+$ satisfy the conservation constraint,
\begin{equation}
  \int (g^+-f) \mathbf{\psi} d \mathbf{u} = \mathbf{0},
\end{equation}
at any location and any time.

Taking the collision time as a local constant, there is an analytic
solution from kinetic model,
\begin{eqnarray}
    f(\mathbf{x},t,\mathbf{u},\xi) &=&
        e^{-t/\tau}f_0(\mathbf{x}-\mathbf{u}t)  \nonumber \\
        && + \frac{1}{\tau}\int_0^{t}g^{+}(\mathbf{x'},t',\mathbf{u},\xi)
        e^{-(t-t')/\tau}dt', \label{eq:localsolution}
\end{eqnarray}
where $ \mathbf{x'} =  \mathbf{x}-\mathbf{u}(t-t')$.



Applying this solution at cell interface,
the mass flux, momentum flux and energy flux can be obtained  as the following,
\begin{eqnarray}
  \mathcal{F}_{macro}=\left(\begin{array}{l} \mathcal{F}_{mass} \\
    \mathcal{F}_{momentum} \\ \mathcal{F}_{energy} \end{array}\right)=
    \left(\begin{array}{l} \mathbf{n}\cdot\int_{\Omega_u}\mathbf{u}fd\mathbf{u} \\
    \mathbf{n}\cdot\int_{\Omega_u}\mathbf{u}\mathbf{u}fd\mathbf{u} \\
    \mathbf{n}\cdot\int_{\Omega_u}\mathbf{u}\frac{1}{2}\mathbf{u}^2 fd\mathbf{u} \end{array}\right).
\end{eqnarray}
The $\Omega_u$ denotes the entire velocity space.


The flux of velocity distribution function at particle velocity $\mathbf{u}_k$
takes the following form:
\begin{eqnarray}
  \mathcal{F}_{\mathbf{u}_k} = \mathbf{n}\cdot\int_{\Omega_{\mathbf{u}_k}}\mathbf{u}fd\mathbf{u},
\end{eqnarray}
where $\Omega_{\mathbf{u}_k}$ denotes the velocity space around $\mathbf{u}_k$.
The first term on the right hand side of Eq.(\ref{eq:localsolution}) can be directly evaluated
from the initial distribution function.
For simplicity, the cell interface is assumed to locate at $\mathbf{x} = 0$.
The normal direction of the cell interface is denoted by $\mathbf{n}$.
Suppose the initial distribution function takes the following form
at a cell interface:
\begin{equation}
  f^n_0(\mathbf{x},\mathbf{u_k},t)|_{t=0} = f^n_{0,k}(\mathbf{x}) =
  \left\{\begin{array}
    {l@{\quad}l} f^L_{0,k}(0)+\frac{\partial f^L}{\partial \mathbf{x}}\cdot \mathbf{x},& x \leq 0, \\
    f^R_{0,k}(0)+\frac{\partial f^R}{\partial \mathbf{x}}\cdot \mathbf{x},& x >0, \end{array} \right.
  \label{eq:f0}
\end{equation}
where nonlinear limiter is used to reconstruct $f^L$, $f^R$, and the corresponding
derivatives.

The second term of Eq.(\ref{eq:localsolution}) corresponds to the
hydrodynamic scale physics which should be constructed from macroscopic quantities.
We can use a continuous distribution function to evaluate the integral term. For
an equilibrium state $g^+$ around a cell interface, it can be formally expressed as following,
\begin{equation}
  g^+(\mathbf{x},\mathbf{u},t) = g^+_0+g^+_{\mathbf{x}}\cdot\mathbf{x}+g^+_t t .
  \label{eq:hydropartofflux}
\end{equation}

In fact, the spatial and temporal derivatives of $g^+$ are the key components for the construction of UGKS.
The derivatives of $g^+$ may be  complicated. Fortunately, only in
the continuum regime this term becomes important.
Here, we use the derivatives of a local Maxwellian distribution to
approximate these quantities. And we have,
\begin{equation}
  g^+(\mathbf{x},\mathbf{u},t) = g^+_0+g_{0,\mathbf{x}}\cdot\mathbf{x}+g_{0,t} t,
  \label{eq:hydropartofflux}
\end{equation}
where $g^+_0$ denotes the post collision state at the beginning of each time step at the cell
interface and $g_0$ is a local Maxwellian distribution function located at $\mathbf{x} = 0$. It can be written as,
\begin{equation}
    g_0 =
        \rho\left(\frac{\lambda}{\pi}\right)^{\frac{K+2}{2}}e^{-\lambda(\mathbf{u-U})^2},
    \label{eq:MaxwellState}
\end{equation}
where $\lambda = {\rho}/{2p}$.


For a specific kinetic model, $g^+_0$ and $g_0$ are uniquely determined
by the initial distribution function $f^n_0$.
For example, the conservative variables
are evaluated by applying the compatibility condition.
The conservation constraint at $(\mathbf{x}=0,t=0+)$ gives
\begin{eqnarray}
  W_0 = \int f_0\psi d\Xi & = & \sum f_{0,k}\psi \nonumber \\
    & = & \sum(f^L_{0,k}H[\mathbf{n}\cdot \mathbf{u}_k]+
    f^R_{0,k}(1-H[\mathbf{n}\cdot \mathbf{u}_k]))\psi,
  \label{eq:conservationConstraint}
\end{eqnarray}
where $H[x]$ is the Heaviside function defined by
\begin{equation}
  H[x] = \left\{\begin{array}{l@{\quad}l}
    0, & x < 0, \\
    1, & x > 0.\end{array}\right.
\end{equation}
Similarly, the high order moments, say, the stress tensor and the
heat flux can be derived by Eq.(\ref{eq:macroMicroRelation}) at a cell interface.
For the details of the numerical reconstruction, please refer to the articles
about gas kinetic scheme \cite{Xu2001,ugks1,ugks1_1}.

Applying the conservation law, the evolution of flow quantities can be obtained.
Owing to the absence of source term, the evolution of the macroscopic
conservative quantities becomes,
\begin{equation}
  W^{n+1} = W^{n}+\frac{1}{V_{\mathbf{x}_i}}\int_{t^n}^{t^{n+1}}\sum_m
    \Delta S_m \mathcal{F}_{macro} dt,
  \label{eq:updateMacro}
\end{equation}
where $V_{\mathbf{x}_i}$ is the volume of $\Omega_{\mathbf{x}_i}$ in the physical space,
$\Delta S_m$ is the area of interface and $m$ is index of surfaces
of $\Omega_{\mathbf{x}_i}$.

The collision term must be considered for the update of the distribution function.
Here, we use two steps to update the distribution function.
\begin{eqnarray}
  f_{\mathbf{u}_k}^{*} & = &
    f_{\mathbf{u}_k}^{n} + \frac{1}{V_{\mathbf{x}_i}}\int_{t^n}^{t^{n+1}}\sum_m
    \Delta S_m \mathcal{F}_{\mathbf{u}_k}+\Delta t \frac{g_{\mathbf{u}_k}^{+(n)}-f_{\mathbf{u}_k}^{n}}{\tau^{n}}, \nonumber\\
  f_{\mathbf{u}_k}^{n+1} & = &
    f_{\mathbf{u}_k}^{n} + \frac{1}{V_{\mathbf{x}_i}}\int_{t^n}^{t^{n+1}}\sum_m
    \Delta S_m \mathcal{F}_{\mathbf{u}_k} \nonumber \\
    & & +\frac{\Delta t}{2}(\frac{g_{\mathbf{u}_k}^{+(*)}-f_{\mathbf{u}_k}^{n+1}}{\tau^{n+1}}
    +\frac{g_{\mathbf{u}_k}^{+(n)}-f_{\mathbf{u}_k}^{n}}{\tau^{n}}),
  \label{eq:updateMicro}
\end{eqnarray}

At first, we derive $f^*_{\mathbf{u}_k}$ as a medium state. And then
solving the second equation, we get $f^{n+1}_{\mathbf{u}_k}$
at the next time level. The above procedure is identical for an arbitrary $g^+$.
For ES-model, $g^{+}$ is written as
\begin{equation}
g^+=\mathcal{G}[f] = \frac{\rho}{\sqrt{\det(2\pi \mathbf{T})}}
\exp(-\frac{1}{2}(\mathbf{u}-\mathbf{U})\cdot \mathbf{T}^{-1}\cdot (\mathbf{u}-\mathbf{U})).
\label{eq:esModel}
\end{equation}
Here, $\mathbf{T}$ is a tensor related to the stress tensor $\mathbf{P}$,
\begin{equation}
\mathbf{T} = (1-C_{es})RT\mathbf{I}+C_{es}\mathbf{P}/\rho,
\end{equation}
where $R$ is gas constant and $T$ is gas temperature.
Andries provided a simple proof that ES-model preserves a correct Prandtl number \cite{Andries2001}.
The same proof can be done for Shakhov model.
In the Shakhov model, the $g^{+}$ takes the form,
\begin{equation}
    g^{+} = \mathcal{M}[f](1+(1-C_{shak}) \mathbf{c} \cdot \mathbf{q}(\frac{\mathbf{c}^2}{RT}-5)/(5pRT)),
    \label{eq:ShakModel}
\end{equation}
where $\mathcal{M}[f]$ denotes the Maxwellian distribution function, $T$ is temperature,
$\mathbf{q}$ is heat flux, $\mathbf{c} = \mathbf{u-U}$
is peculiar velocity and $C_{shak}$ is a parameter which is related to the Prandtl number in this model.


\section{A generalized kinetic model}

Here we discuss the different ways to fix the Prandtl number in the kinetic models.
Following Andries' proof \cite{Andries2001}, we expand the distribution function in continuum regime,
\begin{equation}
f = g^+-\tau(\mathcal{M}[f]_t + \mathbf{u}\cdot\mathcal{M}[f]_{\mathbf{x}})+o(\tau) .
\label{eq:ceExpansion}
\end{equation}
Let's consider the ES-model first. As odd moments of peculiar velocity of
Gaussian function is zero,
the ES-model has no contribution to the heat flux of the distribution function. Therefore,
the Prandtl number is effected only by the variation of stress tensor.
The second term, $-\tau(\mathcal{M}[f]_t+\mathbf{u}\cdot\mathcal{M}[f]_{\mathbf{x}})$, corresponds
to the contribution of BGK model to the stress tensor. The second order moments of
peculiar velocity of Eq.(\ref{eq:ceExpansion}) can be written as the following,
\begin{equation}
\mathbf{P} = (1-C_{es})\rho RT\mathbf{I}+C_{es}\mathbf{P}+O(\tau)_{bgk}.
\end{equation}
Here, the definition of the stress tensor has been considered.
And solving the $\mathbf{P}$, we get,
\begin{equation}
\mathbf{p} = \frac{1}{1-C_{es}}O(\tau)_{bgk}
\end{equation}
Here, $\mathbf{p}$ is the shear stress defined as $\mathbf{p} = \mathbf{P}-\rho RT\mathbf{I}$, and
the $O(\tau)_{bgk}$ corresponds to the shear stress from derivative of local Maxwellian distribution function
which is exactly the shear stress of the BGK model.
The $\mathbf{q}$ from Eq.(\ref{eq:ceExpansion}) will be identical to that in the BGK model. So the Prandtl number of ES-model is,
\begin{equation}
\mbox{Pr} = \frac{1}{1-C_{es}}\mbox{Pr}_{bgk} = \frac{1}{1-C_{es}}.
\end{equation}

For the S-model, according to Eq.(\ref{eq:ShakModel}), the heat flux of distribution function
\begin{equation}
\mathbf{q} = (1-C_{shak})\mathbf{q}+O(\tau)_{bgk},
\end{equation}
where the $O(\tau)_{bgk}$ corresponds to the heat flux from the BGK model. And the $\mathbf{q}$ is
\begin{equation}
\mathbf{q} = \frac{1}{C_{shak}}\mathbf{q}_{bgk}.
\end{equation}
The Shakhov model does not affect the second order moments. So the stress tensor of Shakhov
model keeps unchanged in comparison with the BGK model.
So the Prandtl number for Shakhov model is,
\begin{equation}
\mbox{Pr}_{shak} = C_{shak}\mbox{Pr}_{bgk} = C_{shak}.
\end{equation}

These proofs imply that it is sufficient to achieve a correct continuum limit
as long as the spatial and temporal derivatives are expressed as the expansion of
local Maxwellian.
So, it is appropriate for the hydrodynamic flux
to be estimated by the derivative of the Maxwellian function
in the integral solution as mentioned
in the last section.
Furthermore, the proofs also show that the kinetic models fix the Prandtl
number via the adjustment of either stress or
heat flux of the relaxation term.
It is quite straightforward to combine these two approaches together.

The ES-model and S-model change either
the stress tensor or the heat flux of the post collision terms to achieve
a correct Prandtl number. How about to change these two quantities simultaneously.
It's obvious that this kind of modification could also generate a correct Prandtl
number, and provides a free parameter as a by-product.

Specifically, the post collision term of the generalized kinetic model is
\begin{equation}
    g^{+} = \mathcal{G}[f]+\mathcal{S}[f],
\end{equation}
where the $\mathcal{G}[f]$ is defined by Eq.(\ref{eq:esModel}).
The $\mathcal{S}[f]$ is Eq.(\ref{eq:ShakModel}) for Shakhov model
without the first equilibrium state, namely,
\begin{equation}
    \mathcal{S}[f] = \mathcal{M}[f][(1-C_{shak}) \mathbf{c} \cdot \mathbf{q}(\frac{c^2}{RT}-5)/(5pRT)].
\end{equation}

The two coefficients, $C_{es}$ and $C_{shak}$, are two independent parameters at this moment.
In order to obtain the right transport coefficients, we still follow the
proof of Andries \cite{Andries2001}.
Eq. (\ref{eq:ceExpansion}) changes to the following one,
\begin{equation}
f = \mathcal{G}[f]+\mathcal{S}[f]-\tau(\mathcal{M}_t+\mathbf{u}\cdot\mathcal{M}_{\mathbf{x}})+o(\tau).
\end{equation}
For stress tensor,
\begin{equation}
\mathbf{P} = (1-C_{es})\rho RT\mathbf{I}+C_{es}\mathbf{P}+\mathbf{p}_{bgk},
\end{equation}
then
\begin{equation}
\mathbf{p} = \frac{1}{(1-C_{es})}\mathbf{p}_{bgk}.
\end{equation}
And for heat flux,
\begin{equation}
\mathbf{q} = (1-C_{shak})\mathbf{q}+\mathbf{q}_{bgk},
\end{equation}
then
\begin{equation}
\mathbf{q} = \frac{1}{C_{shak}}\mathbf{q}_{bgk}.
\end{equation}
As a result, the Prandtl number for the generalized kinetic model is
\begin{equation}
\mbox{Pr} = \frac{C_{shak}}{1-C_{es}}\mbox{Pr}_{bgk} = \frac{C_{shak}}{1-C_{es}}.
\end{equation}
And the viscosity is
\begin{equation}
\mu = \frac{\tau p}{1-C_{es}}.
\end{equation}
If Prandtl number is fixed, there is a free parameter in the generalized model.
Here, the $C_{es}$ can be taken as the free parameter. When $C_{es} = 0$ and $C_{shak} = \mbox{Pr}$ , the
generalized model is identical with the Shakhov model. When
$C_{es} = 1-\frac{1}{\mbox{Pr}}$ and $C_{shak} = 1$,
it gives the ES-model.
When $C_{es} = 0$ and $C_{shak} = 1$, it presents the BGK model.
And for the other values, the generalized kinetic model
shows how the ES-model changes to Shakhov model continuously.
And the new free parameter might provide an opportunity to preserve additional
physical properties in the full Boltzmann collision term.

The generalized kinetic model is employed in the UGKS introduced in section \ref{sec:scheme} for its numerical solution.

\section{Numerical results}
\label{sec:results}

\subsection{Shock structure}
The shock structure is a typical example of non-equilibrium flow structure, and is
a distinguishable test case. Kinetic models
show very different performances in shock structure simulation.
To examine capabilities of ES-model and S-model, Mach 8 argon
shock structure is simulated, and the solutions are
compared with the DSMC results \cite{Bird94}. The DSMC code is provided by G.A. Bird.
The viscosity-temperature coefficient $\omega$ is 0.81, namely, $\mu \sim T^{0.81}$.
The Prandtl number is $2/3$, and $C_{es}$ varies from $-0.5$ to $0.5$ in the generalized kinetic model.
The two special cases, the ES-model and the S-model, are included in this set of simulations.
The reference viscosity is determined as following,
\begin{eqnarray}
\mu_{ref} = \frac{30}{(7-2\omega)(5-2\omega)}\frac{\rho\lambda\sqrt{2\pi RT}}{4} .
\label{eq:refvis}
\end{eqnarray}
In our simulation, the spatial coordinate is normalized by the upstream mean free path, namely,
the upstream mean free path of argon is 1.
The computational spatial domain is $[-50,30]$ and is uniform meshed by 300 grids.
\begin{figure}[h]
    \parbox[t]{0.48\textwidth}{
    \includegraphics[totalheight=6cm, bb = 90 35 690 575, clip =
    true]{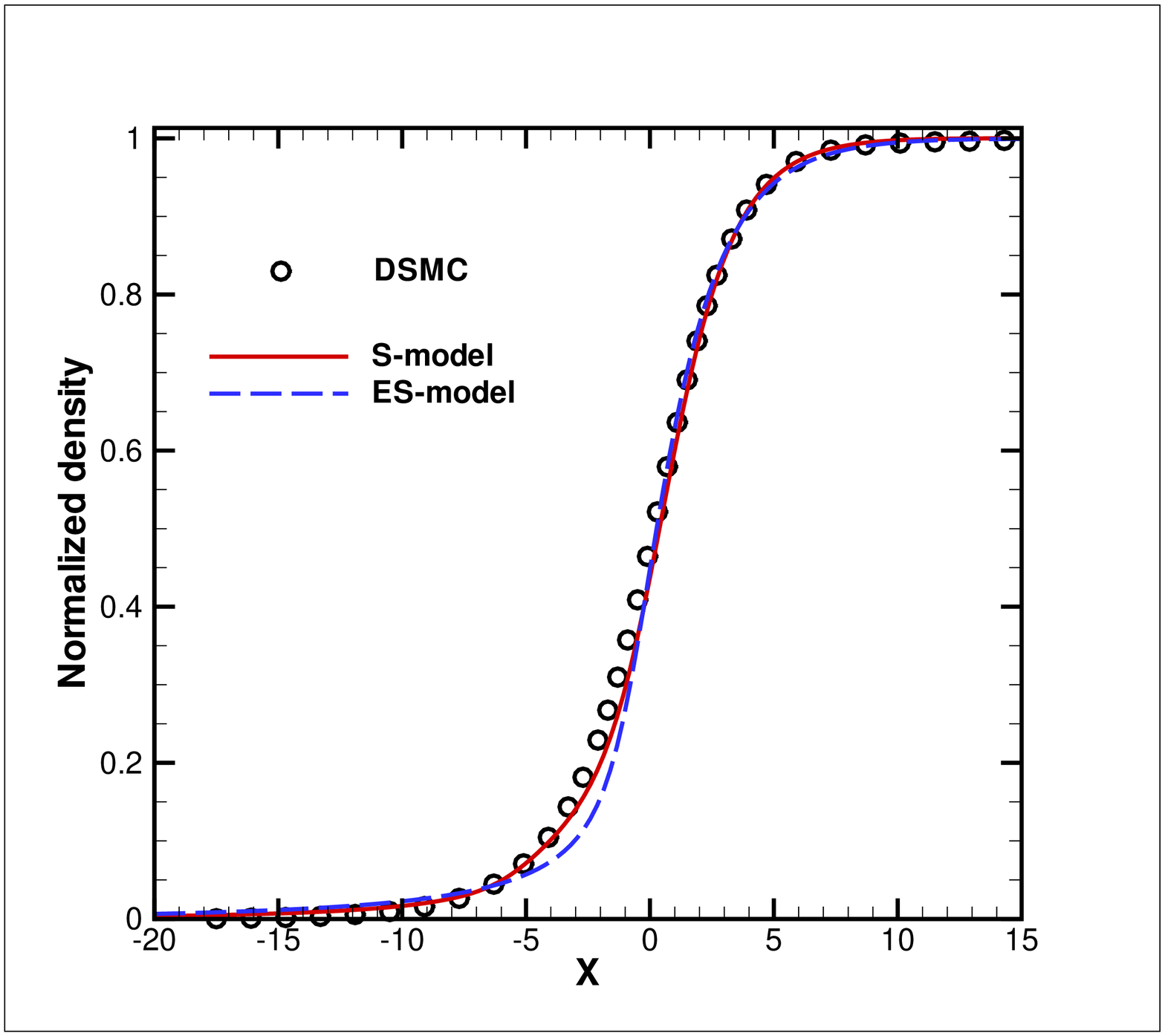}
    }
    \hfill
    \parbox[t]{0.48\textwidth}{
    \includegraphics[totalheight=6cm, bb = 90 35 690 575, clip =
    true]{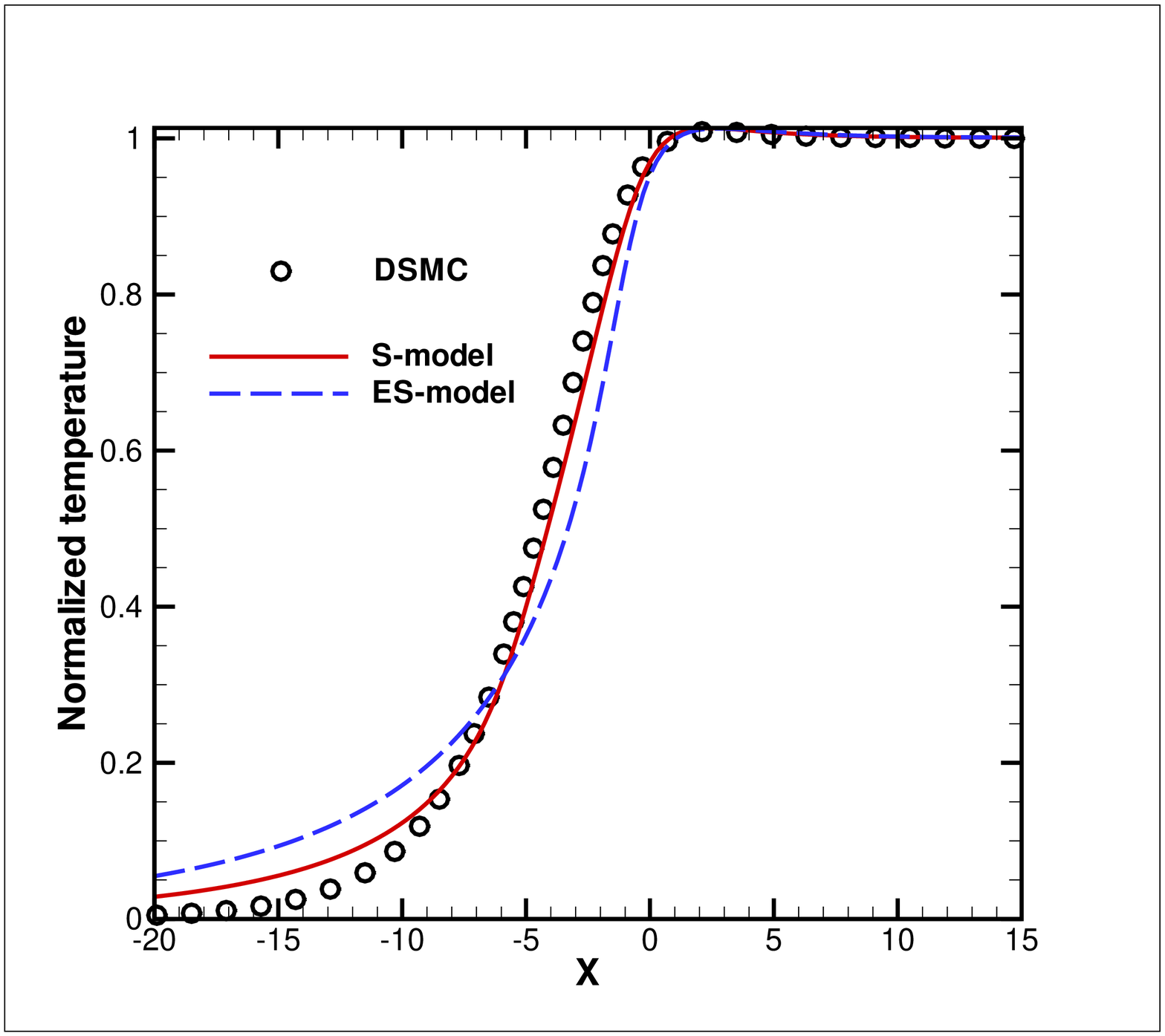}
    }
    \caption{The shock structure for ES-model and S-model at $\mbox{Ma} = 8$ and $\omega = 0.81$.}
    \label{fig:shock1} 
\end{figure}

\begin{figure}[h]
    \parbox[t]{0.48\textwidth}{
    \includegraphics[totalheight=6cm, bb = 90 35 690 575, clip =
    true]{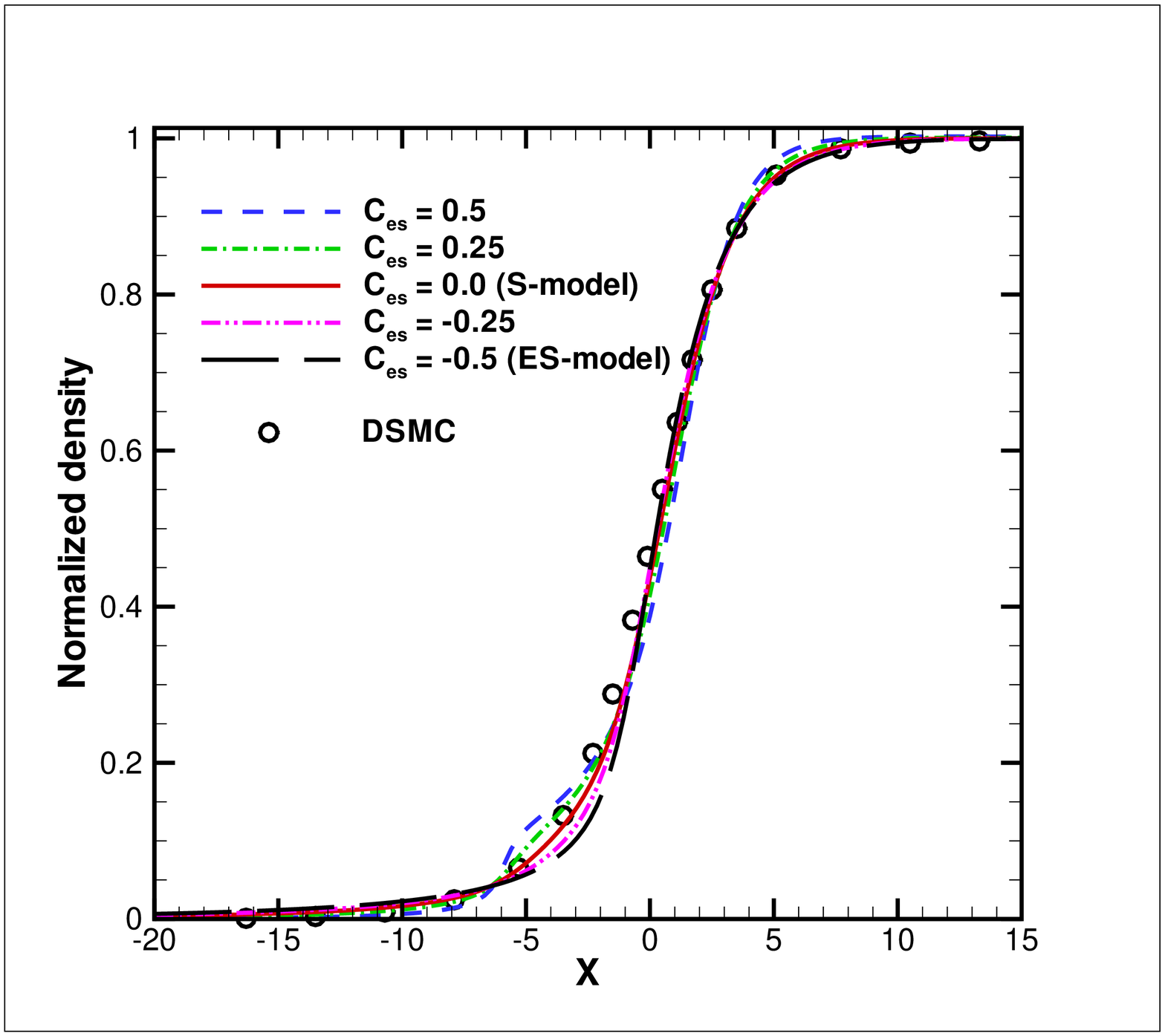}
    }
    \hfill
    \parbox[t]{0.48\textwidth}{
    \includegraphics[totalheight=6cm, bb = 90 35 690 575, clip =
    true]{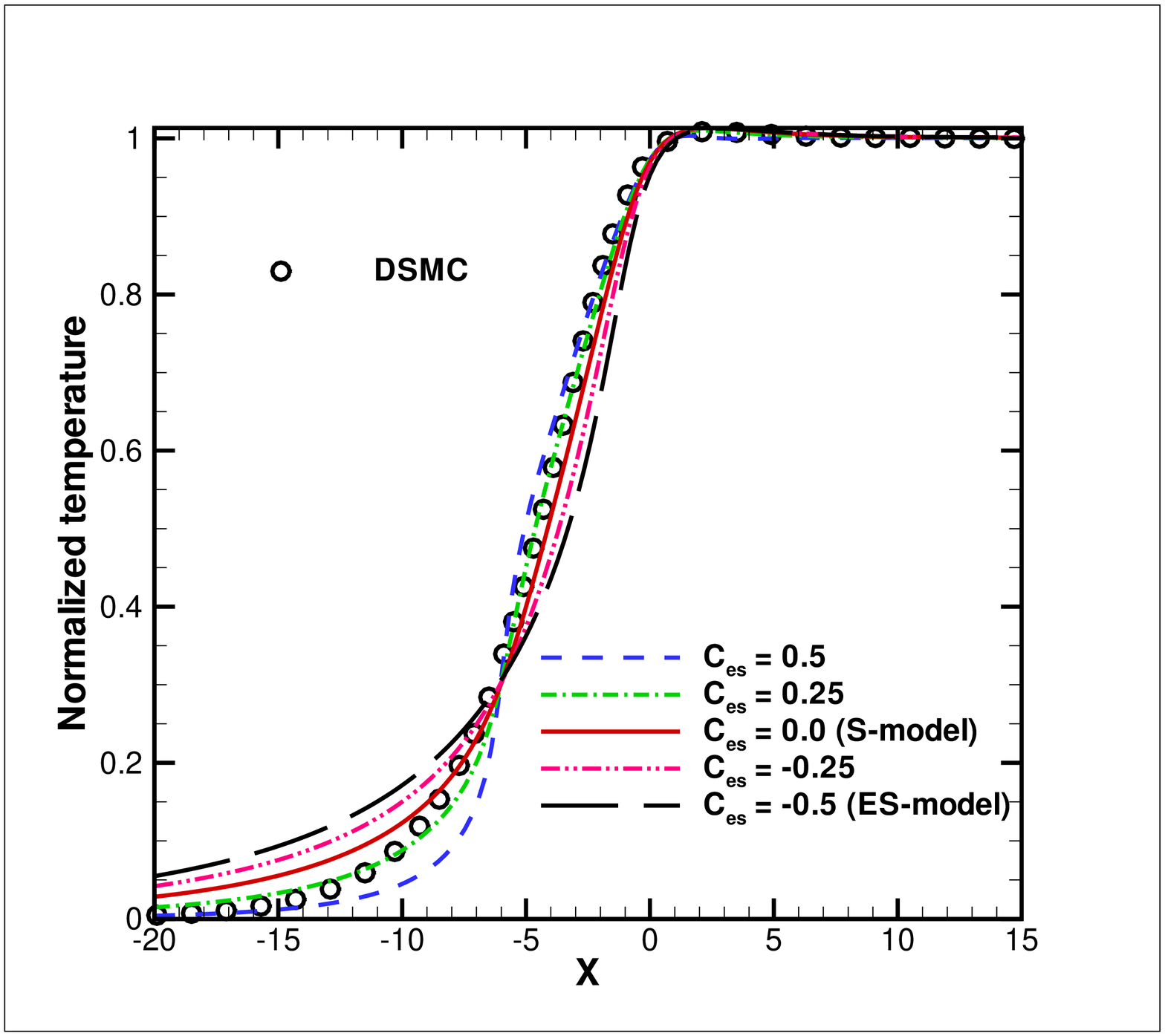}
    }
    \caption{The shock structure for the generalized kinetic model with different $C_{es}$ at $\mbox{Ma} = 8$, and $\omega = 0.81$.}
    \label{fig:shock2} 
\end{figure}

Figure \ref{fig:shock1} gives the density and temperature profiles from the
ES-model and S-model. The S-model and the DSMC present almost identical density profiles.
But the temperature rises a little bit early for the S-model. In comparison with S-model, the ES-model
predicts a narrow density profile and a wide temperature profile. Obviously, the S-model
performs much better than ES-model in this case.

Figure \ref{fig:shock2} shows the tendency how the shock structure changes
while the $C_{es}$ varies from $-0.5$ to $0.5$. Note that $\mbox{Pr}=2/3$ is fixed in all these results.
The generalized kinetic model presents a set of shock structures with the same Prandtl number.
When the $C_{es} = -0.5$, the generalized kinetic model presents the ES-model. As the value of $C_{es}$
becomes larger, the temperature profile becomes steeper.
Meanwhile, the density profile grows wider.
When $C_{es}$ is larger than 0, the temperature profile still becomes steepening.
But when $C_{es}$ exceeds $0.15$, the density profile turns out to be twisted near the upstream.
Although the annoyed twisting density profile makes this range of the free parameter unacceptable,
the strong dependence of the $C_{es}$ is confirmed. This coefficient effects the behavior of
kinetic model.

Taking moments of the generalized kinetic model, consider the following three equations for different moments,
\begin{eqnarray}
\frac{\partial f}{\partial t} &=& -\frac{1}{\tau}(f-g^+), \\
\frac{\partial P_{ij}}{\partial t} = \frac{\partial p_{ij}}{\partial t} &=& \frac{-(1-C_{es})}{\tau}p_{ij}, \\
\frac{\partial q_{i}}{\partial t} &=& \frac{-C_{shak}}{\tau}q_{i},
\end{eqnarray}
which determine three different relaxation processes,
namely, the relaxation of distribution function itself, the relaxation of second order
moments and the relaxation of third order moments.
The ratios between different relaxation rates are determined by the two coefficients,
$C_{es}$ and $C_{shak}$.
Let the Prandtl number fixed, i.e., $\mbox{Pr}={C_{shak}}/{(1-C_{es})}$ keeps constant
when changing $C_{es}$. The $(1-C_{es})$ gives the ratio
between the relaxation of distribution function
and the relaxation of the second moments of distribution function.
This is the physical meaning of the $C_{es}$.
For example, if $(1-C_{es})$ is bigger than 1, the second order moments decease more rapidly
than the distribution function itself.
As shown in figure \ref{fig:shock2}, different $C_{es}$ presents different relaxation ratio and
provides different shock structures.

\subsection{Force driven Poiseuille flow}
\begin{figure}[h]
    \parbox[t]{0.48\textwidth}{
    \includegraphics[totalheight=6cm, bb = 90 35 690 575, clip =
    true]{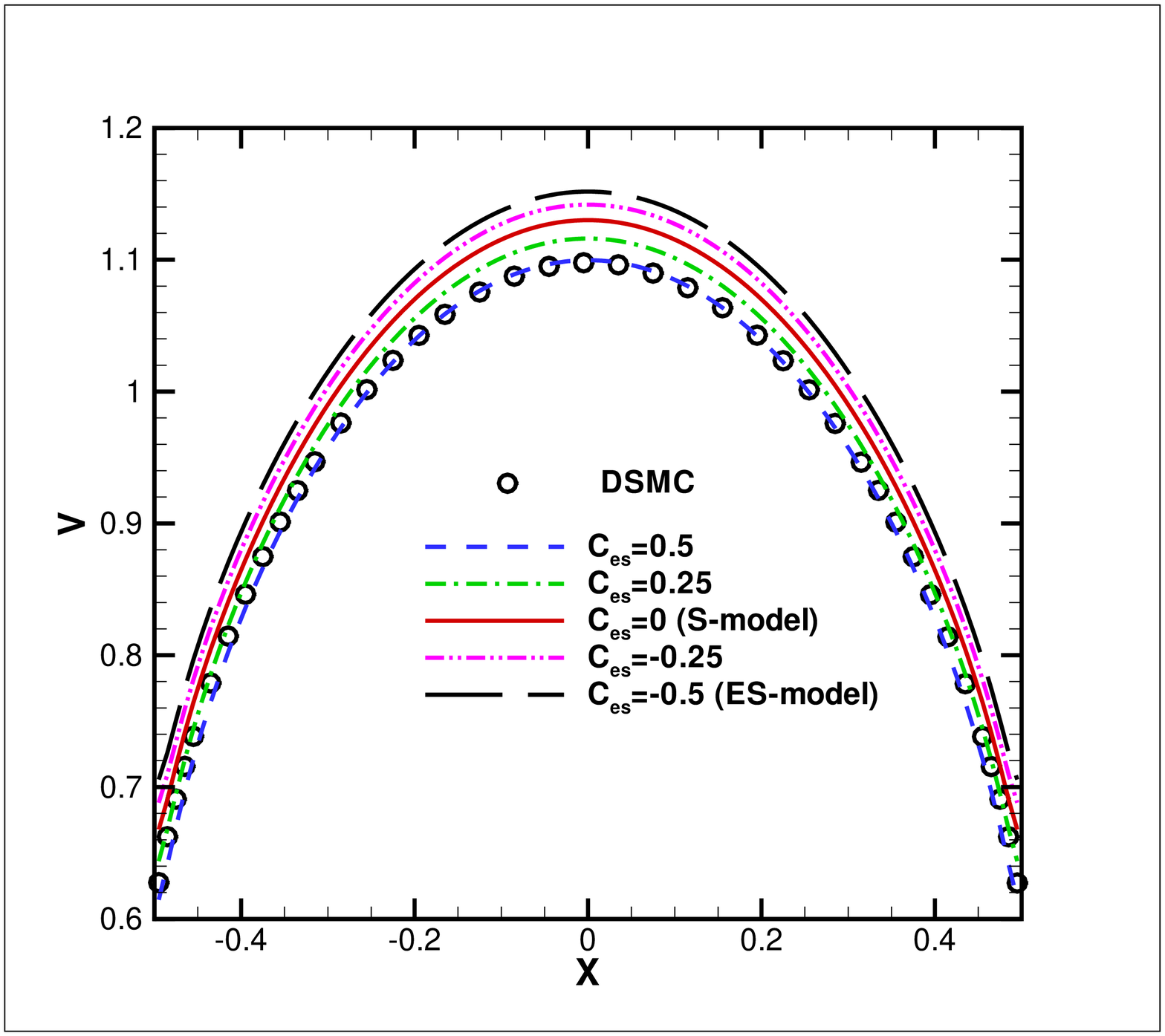}
    }
    \hfill
    \parbox[t]{0.48\textwidth}{
    \includegraphics[totalheight=6cm, bb = 90 35 690 575, clip =
    true]{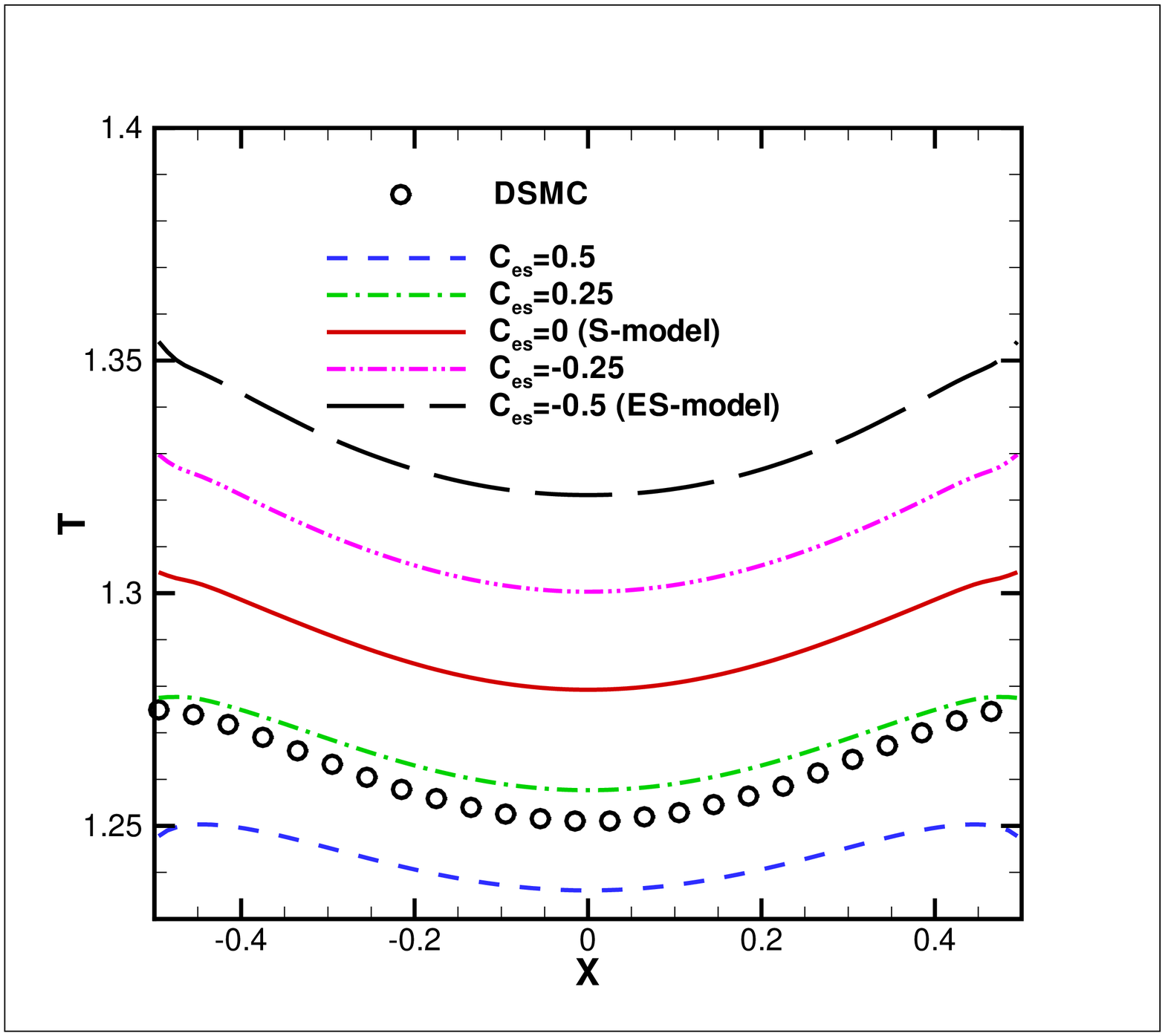}
    }
    \parbox[t]{0.48\textwidth}{
    \includegraphics[totalheight=6cm, bb = 90 35 690 575, clip =
    true]{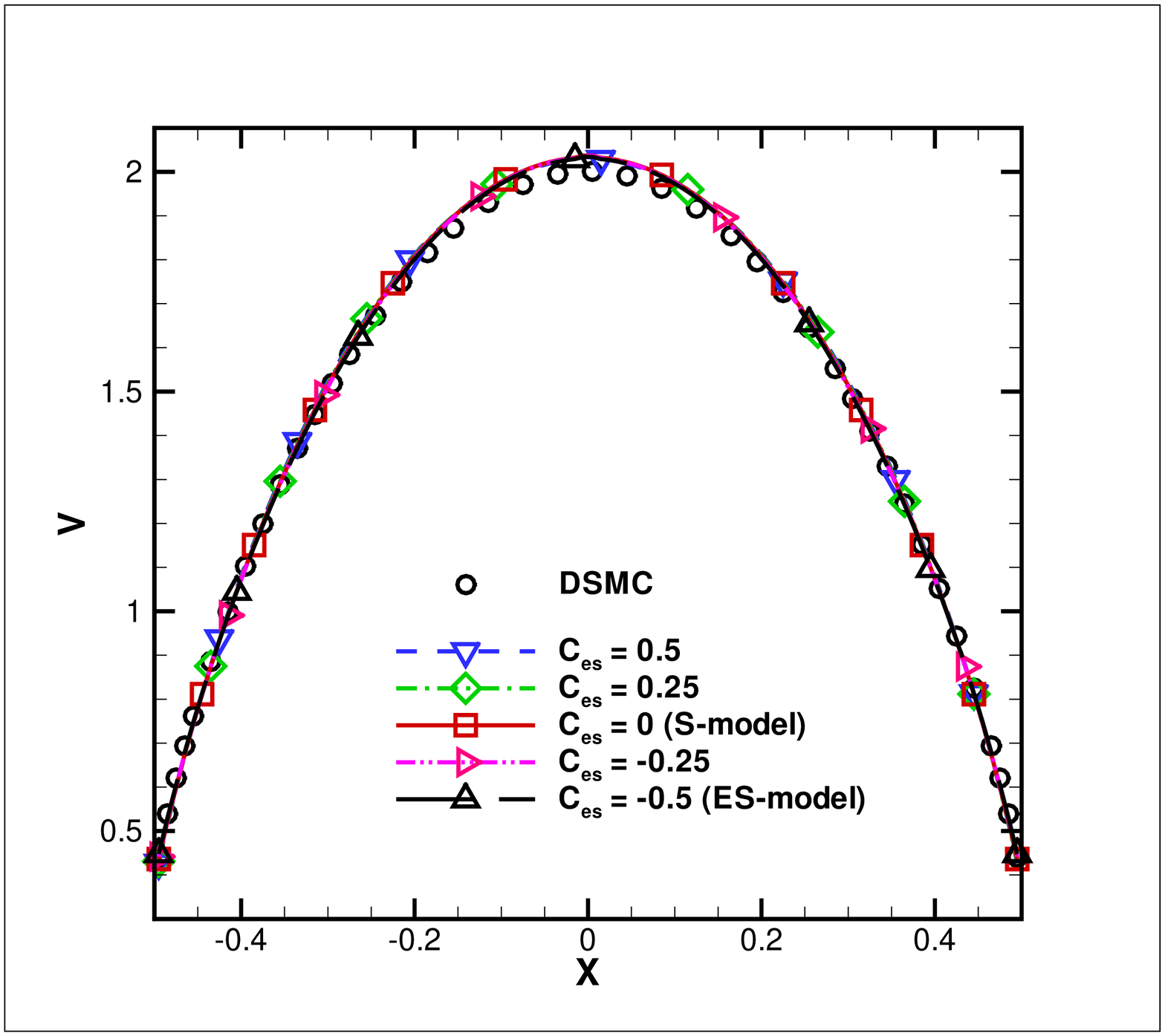}
    }
    \hfill
    \parbox[t]{0.48\textwidth}{
    \includegraphics[totalheight=6cm, bb = 90 35 690 575, clip =
    true]{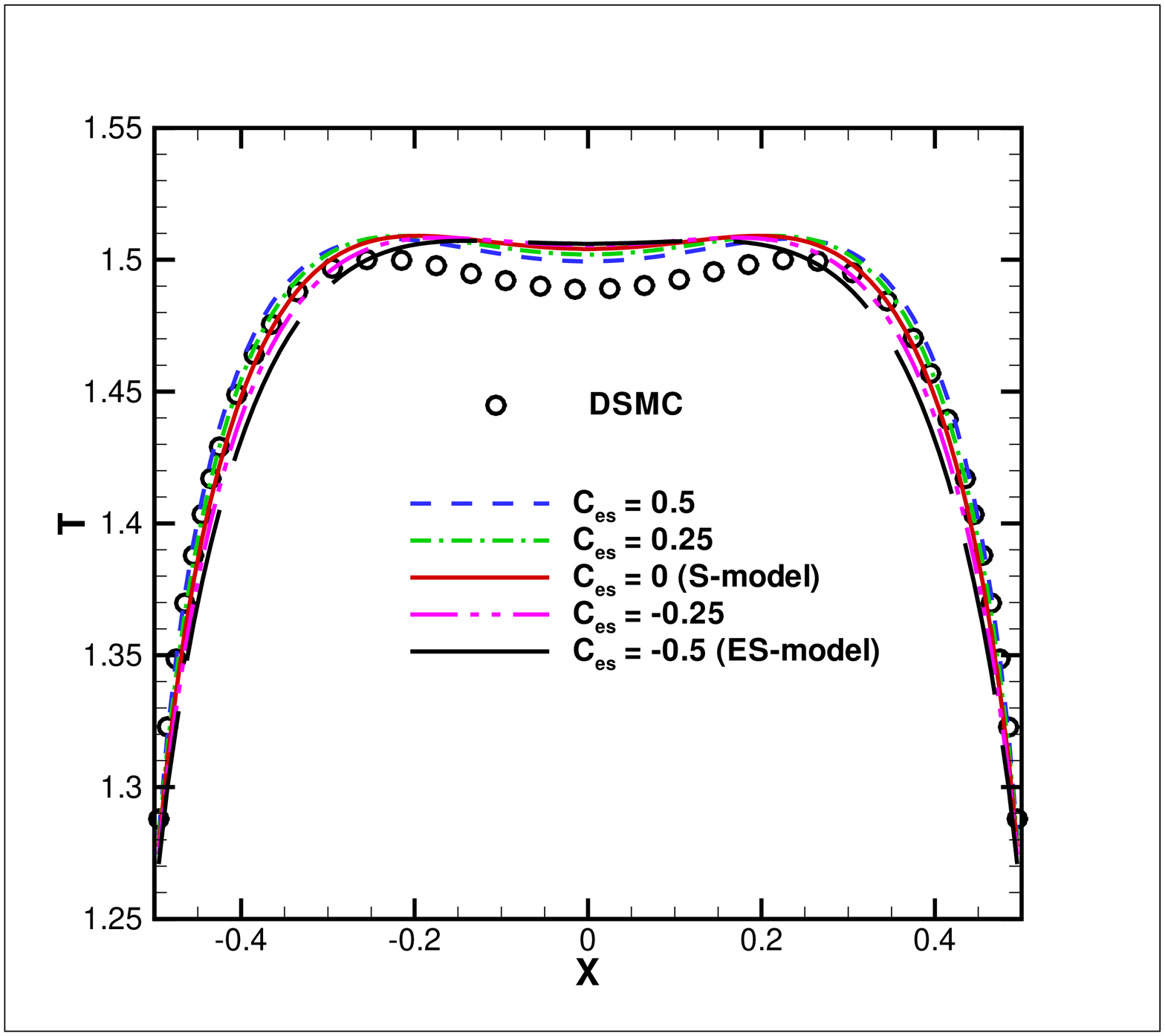}
    }
    \parbox[t]{0.48\textwidth}{
    \includegraphics[totalheight=6cm, bb = 90 35 690 575, clip =
    true]{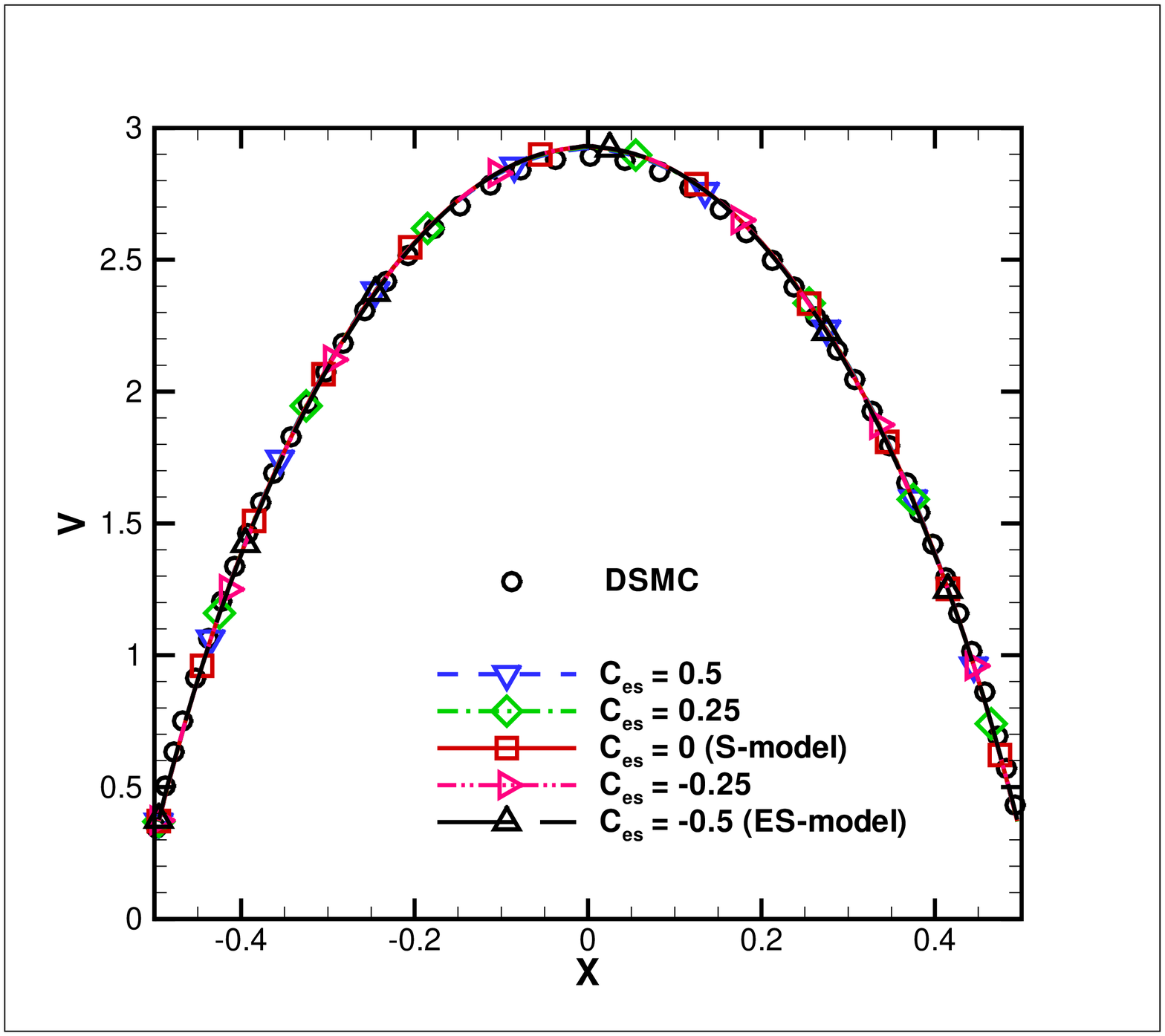}
    }
    \hfill
    \parbox[t]{0.48\textwidth}{
    \includegraphics[totalheight=6cm, bb = 90 35 690 575, clip =
    true]{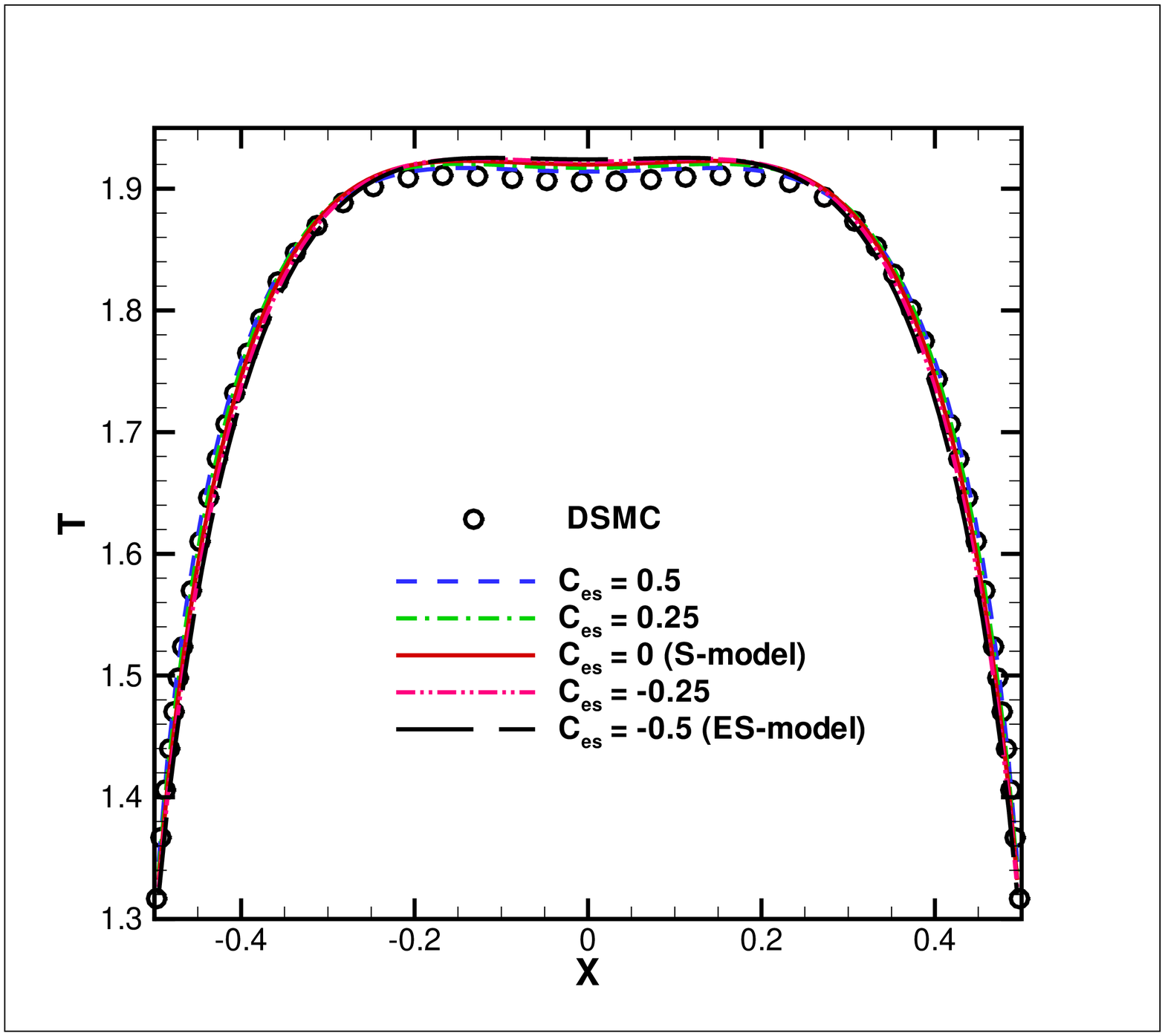}
    }
    \caption{The velocity and temperature profiles from the generalized kinetic model under different Knudsen numbers.
    The Knudsen numbers are 1, 0.1 and 0.05 respectively from top to the bottom. And the Gravity is $G = 1$}
    \label{fig:poiseuille2} 
\end{figure}

In the force driven Poiseuille flow, the external force drives the flow motion between
two fixed plates. The flow field will achieve a steady state when the external
force is balanced by the shear stress from the fixed boundaries.
We also consider monatomic gas in this simulation. To follow the study in \cite{Meng2013},
the Knudsen number is defined as,
\begin{eqnarray}
\mbox{Kn} = \sqrt{\frac{\pi}{2}}\frac{\mu_0\sqrt{RT_0}}{p_0 L},
\end{eqnarray}
where $L$ is the width of the channel, and subscript $0$ denotes the initial value
of variable. The gas is confined between two vertical plates which locate at $x = -0.5$ and
$x = 0.5$ respectively. The temperature of the plates is $T_w = 1$. The initial flow states
are shown as following, $T_0 = 1$, $\rho_0 = 1$, $p_0 = 1$. The gravity is represented by $G$,
and is in the vertical direction. Here the hard sphere molecule is adopted, namely,
the viscosity-temperature coefficient is $\omega = 0.5$.
The gas-wall interaction uses fully diffusive kinetic boundary condition.
Due to the large value of $G=1$, this test can become a very tough one and the distribution function
is fully distorted by the external forcing, especially at high Knudsen number.

As shown in figure \ref{fig:poiseuille2}, when the Knudsen number is small, say, $\mbox{Kn} = 0.05$,
the difference between results from different kinetic models and the DSMC is small. But, as the Knudsen
number becomes large, the temperature profiles separate from each other. Similar to the shock structure,
for all Knudsen number, the results from S-model are closer to the DSMC resluts than the ES-model.
The profiles with different $C_{es}$ cover the results of ES-model and S-model. And when the
$C_{es}$ is larger than $0$, the temperature profile moves from the S-model result to the DSMC
result. It is clearly shown
that the generalized kinetic model can predict more accurate  results in comparison with the S-model and ES-model if
$C_{es}$ is specified properly.

\subsection{Unsteady boundary heating}
\begin{figure}[h]
    \centering
    \parbox[t]{1.0\textwidth}{
    \centering
    \mbox{$(a) \mbox{Kn} = 0.1,\ \theta = \phi$}
    }
    \parbox[t]{0.48\textwidth}{
    \includegraphics[totalheight=5.5cm, bb = 90 35 690 575, clip =
    true]{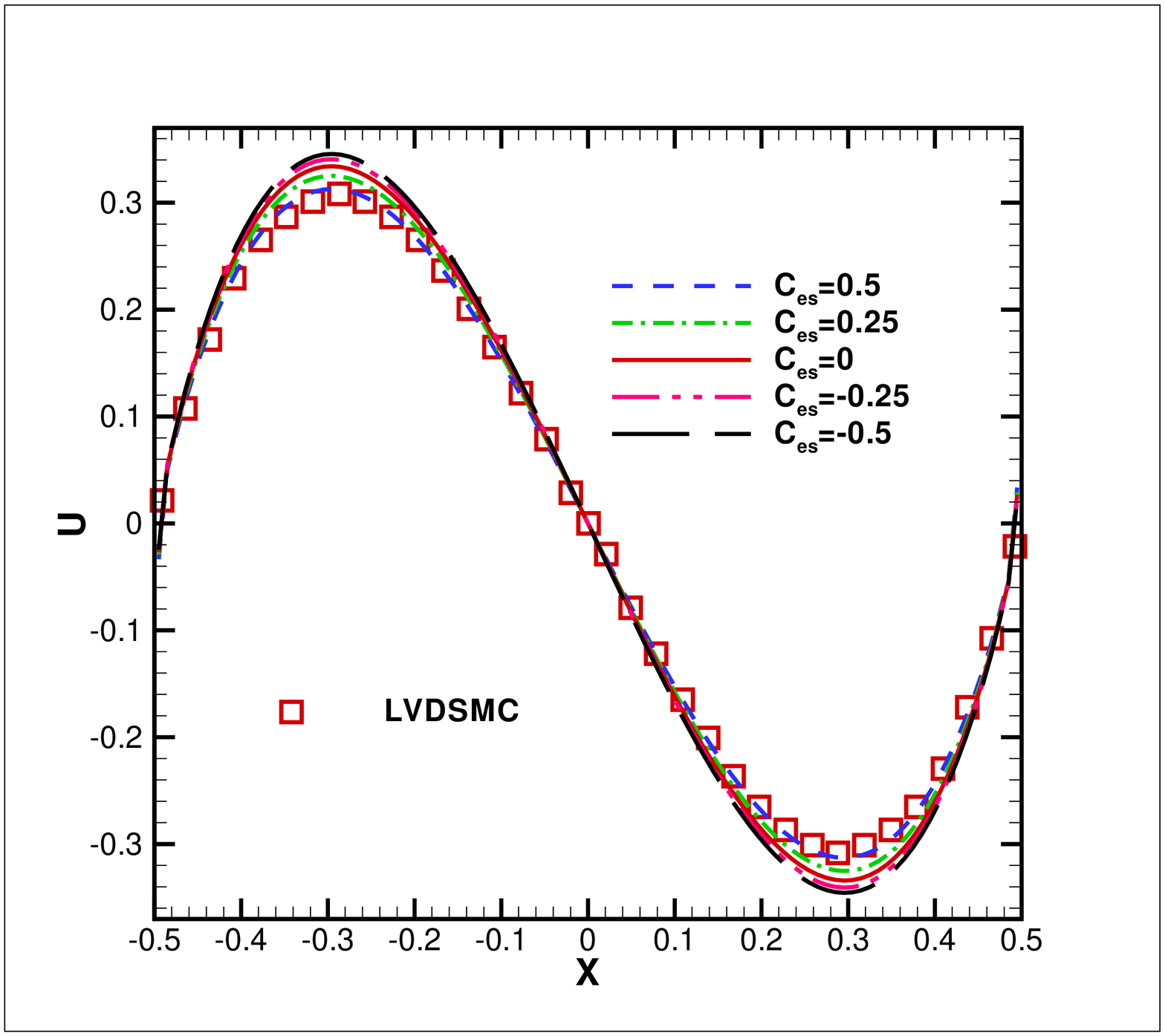}
    }
    \hfill
    \parbox[b]{0.48\textwidth}{
    \includegraphics[totalheight=5.5cm, bb = 90 35 690 575, clip =
    true]{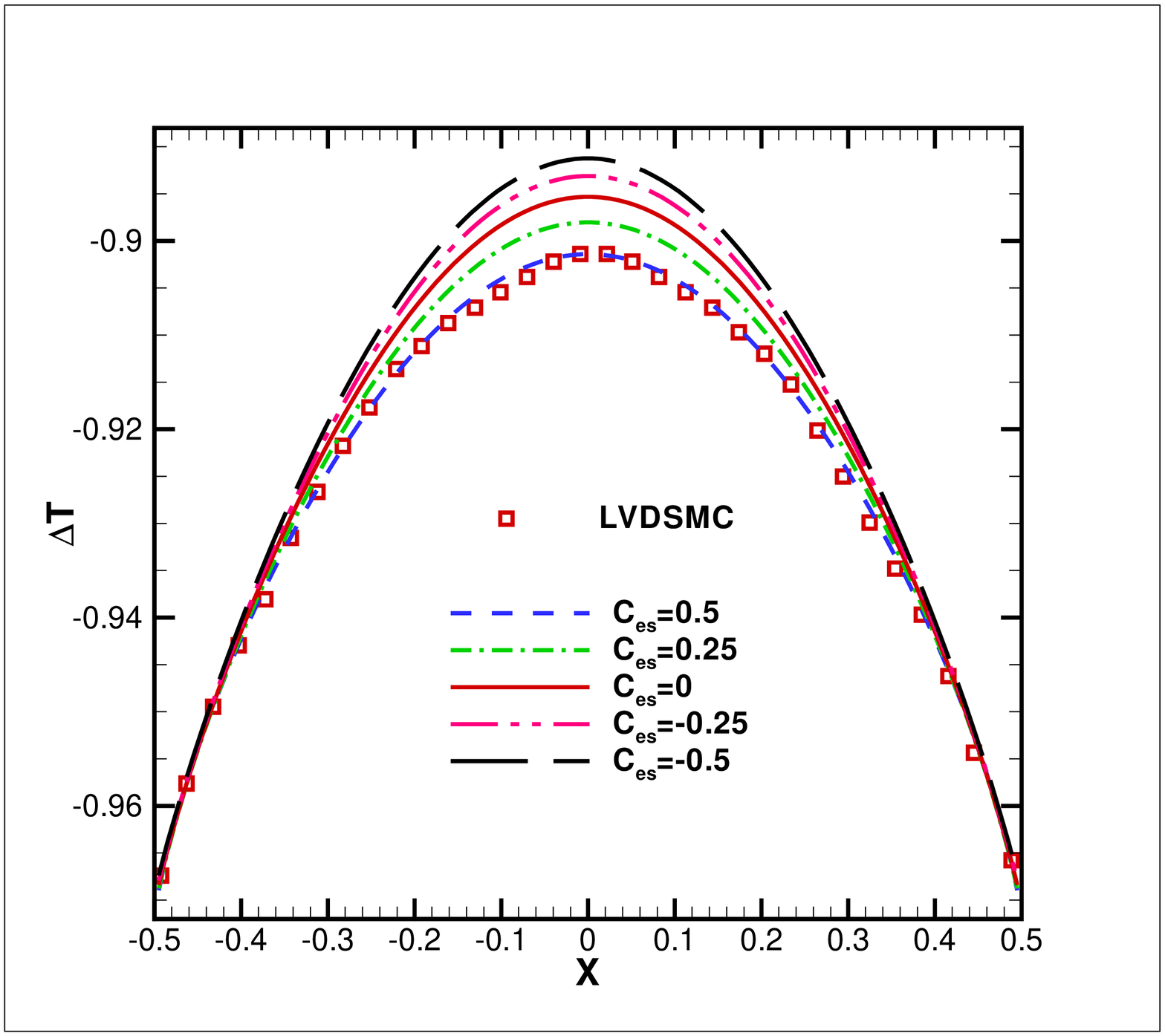}
    }
    \centering
    \parbox[t]{1.0\textwidth}{
    \centering
    \mbox{$(b) \mbox{Kn} = 0.2,\ \theta = 4\phi$}
    }
    \parbox[t]{0.48\textwidth}{
    \includegraphics[totalheight=5.5cm, bb = 90 35 690 575, clip =
    true]{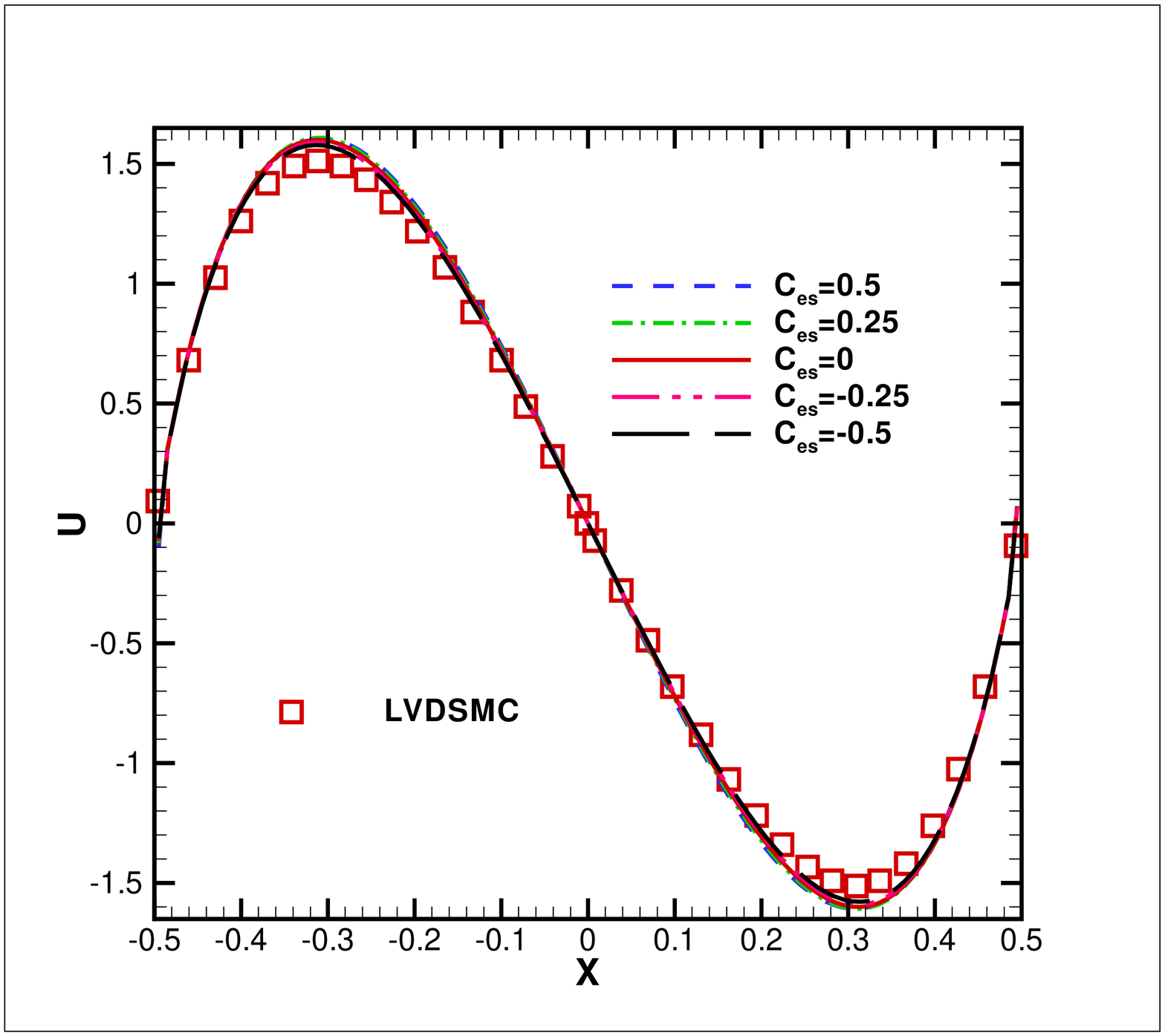}
    }
    \hfill
    \parbox[b]{0.48\textwidth}{
    \includegraphics[totalheight=5.5cm, bb = 90 35 690 575, clip =
    true]{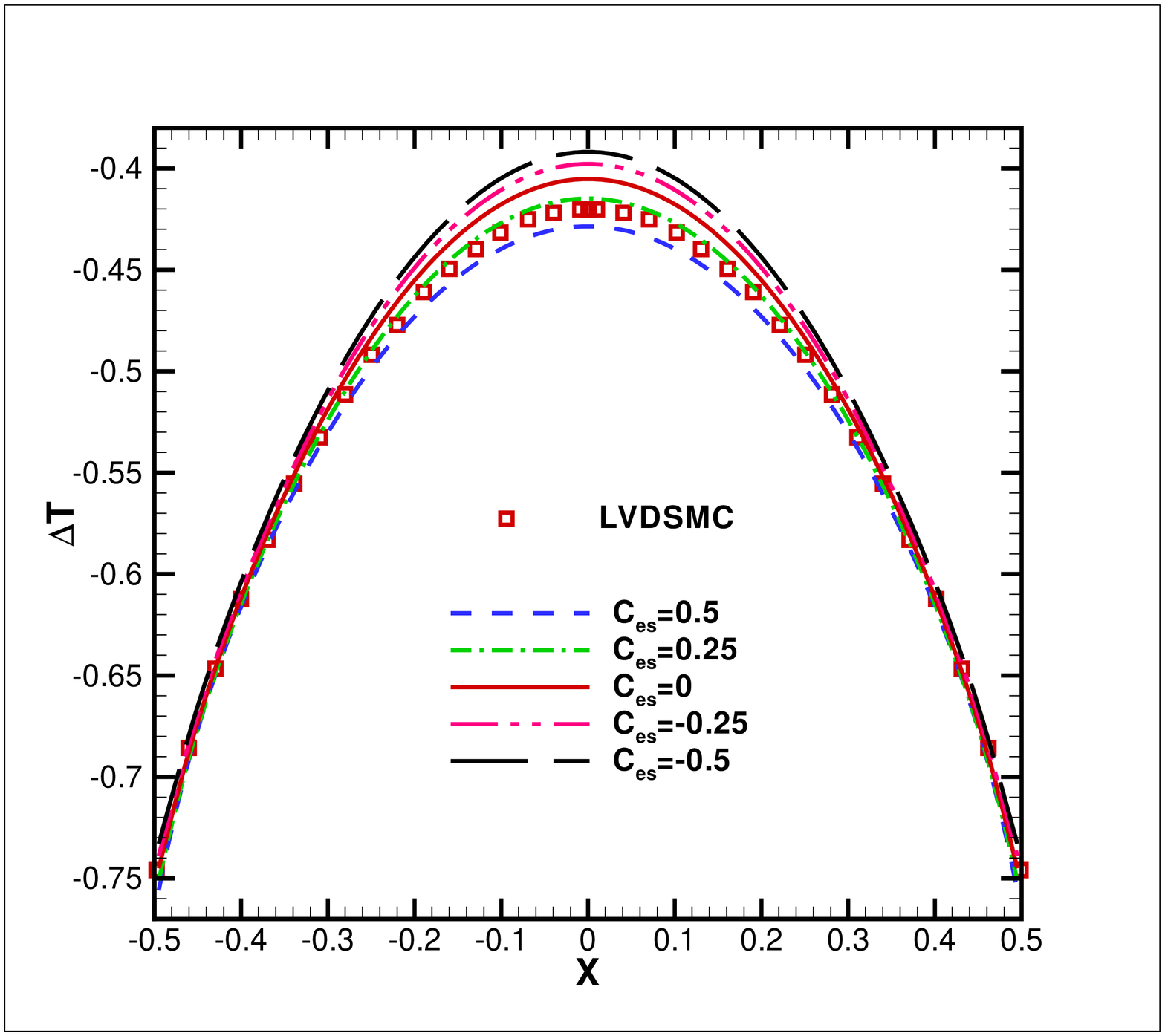}
    }
    \centering
    \parbox[t]{1.0\textwidth}{
    \centering
    \mbox{$(c) \mbox{Kn} = 0.5,\ \theta = 4\phi$}
    }
    \parbox[t]{0.48\textwidth}{
    \includegraphics[totalheight=5.5cm, bb = 90 35 690 575, clip =
    true]{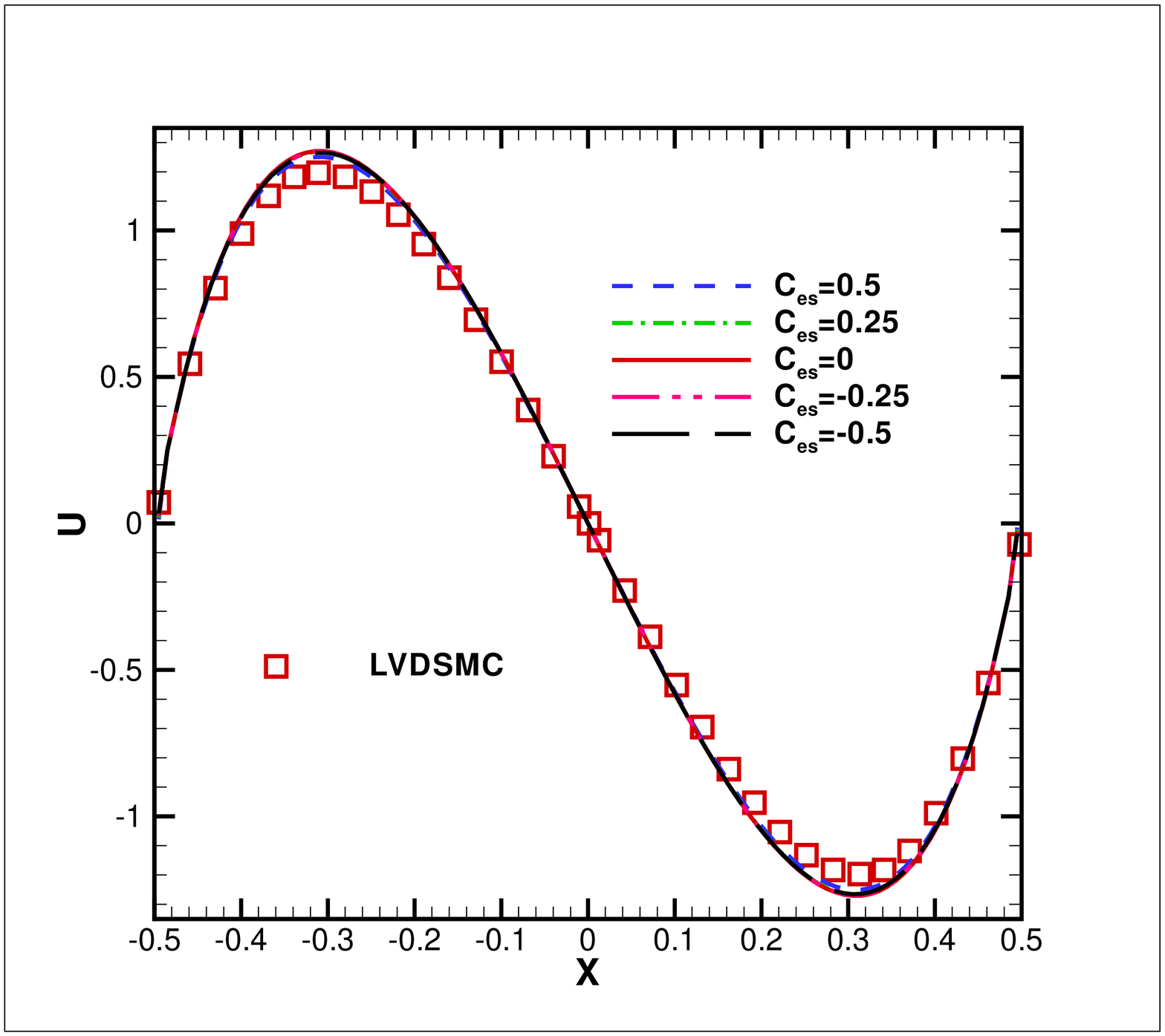}
    }
    \hfill
    \parbox[b]{0.48\textwidth}{
    \includegraphics[totalheight=5.5cm, bb = 90 35 690 575, clip =
    true]{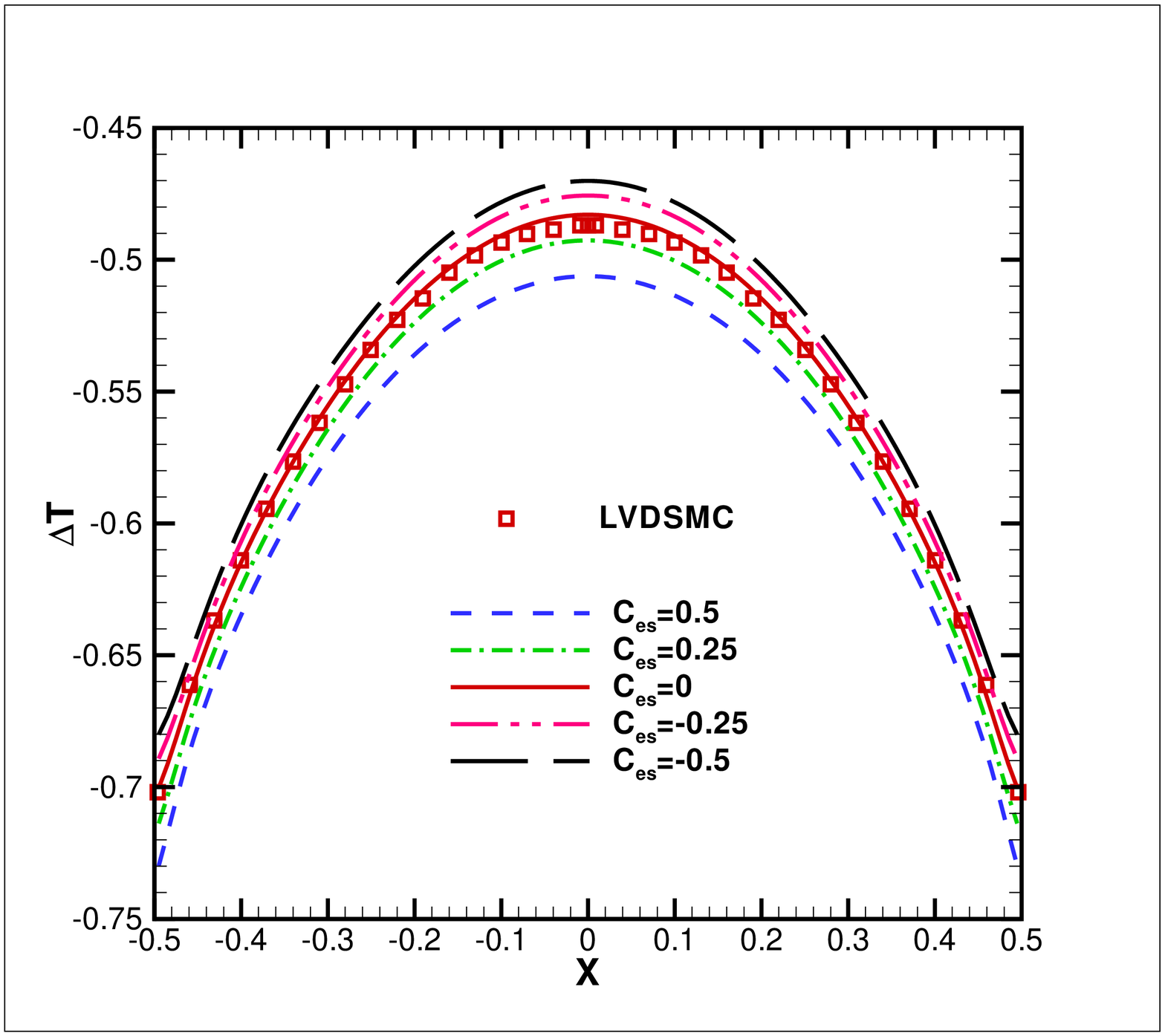}
    }
    \caption{The velocity and temperature profile for unsteady boundary heating problem
    at $\theta t = 3\pi/2$. The $\phi$ is defined as $\phi = \pi\sqrt{2}/16$. }
    \label{fig:UBH} 
\end{figure}
In this section we solve the unsteady flow problem using the generalized kinetic model. The numerical configuration
is identical to the unsteady boundary heating problem in reference \cite{Meng2013jfm}.
The gas is heated by two wall with time-dependent temperatures $T_w = 1 + 0.002\sin(\theta t)$.
Hard sphere molecule is adopted in the simulation. And the Prandtl number is $2/3$. The Knudsen number is
defined as following,
\begin{eqnarray}
\mbox{Kn} = \frac{16}{5\sqrt{2\pi}}\frac{\mu_0\sqrt{RT_0}}{p_0 L}.
\end{eqnarray}
Figure \ref{fig:UBH} presents
the velocity and temperature profiles at $\theta t = 3\pi/2$. The $U$ velocity is normalized by $2\times 10^{-5}$,
and $\Delta T$ is defined as $\Delta T = (T-T_0)/0.002$. Obviously, the results from
S-model is closer to the LVDSMC results. When $C_{es}$ is larger than 0, the generalized kinetic model
gives a better result. This coefficient is very close to the one in the force driven Poiseuille flow.

\subsection{Response of a gas to a spatially varying boundary temperature}
\begin{figure}[h]
    \parbox[t]{0.45\textwidth}{
    \centering
    \mbox{S-model}
    \includegraphics[totalheight=6cm, bb = 141 82 583 503, clip =
    true]{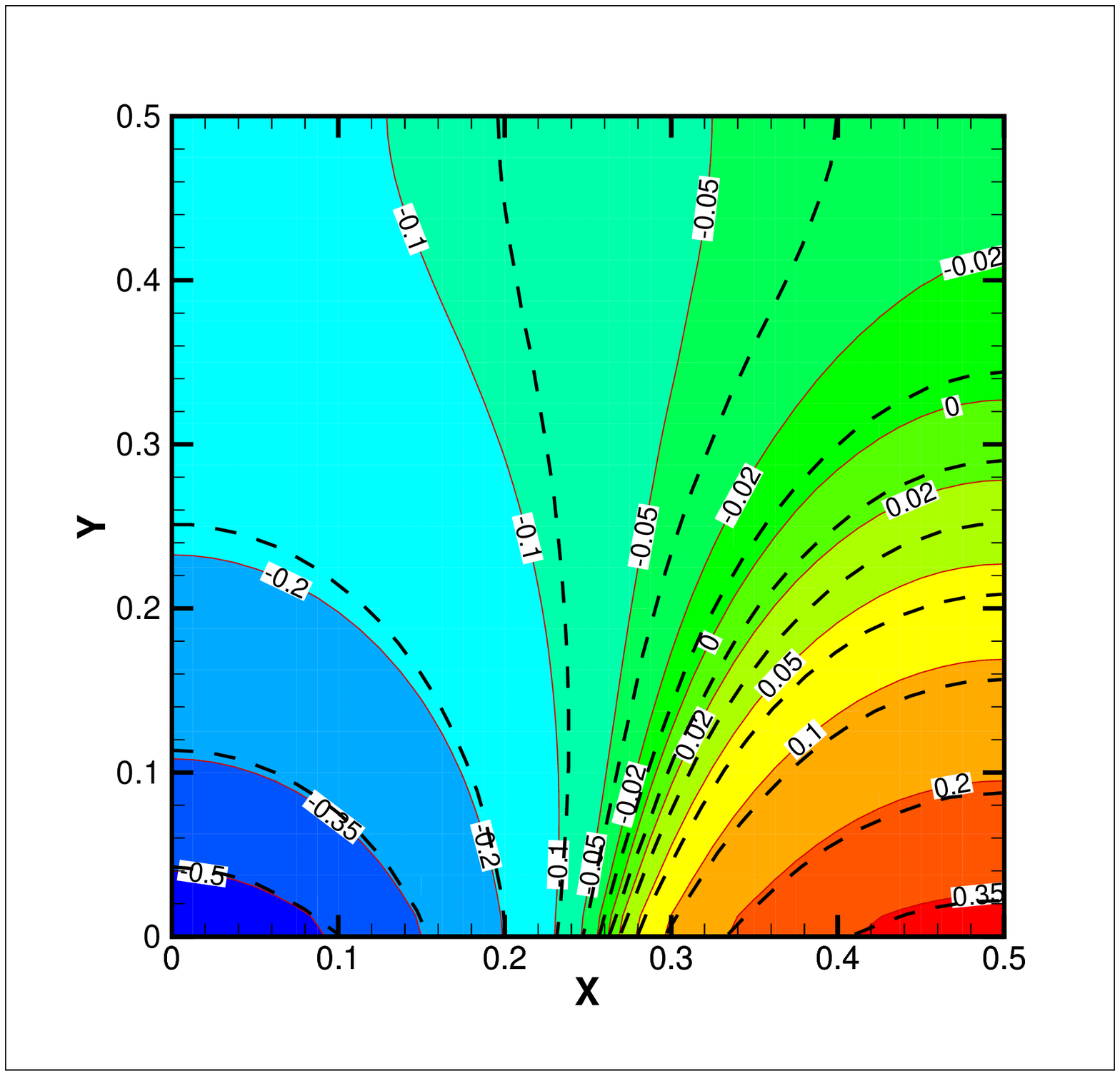}
    }
    \hfill
    \parbox[t]{0.45\textwidth}{
    \centering
    \mbox{ES-model}
    \includegraphics[totalheight=6cm, bb = 141 82 583 503, clip =
    true]{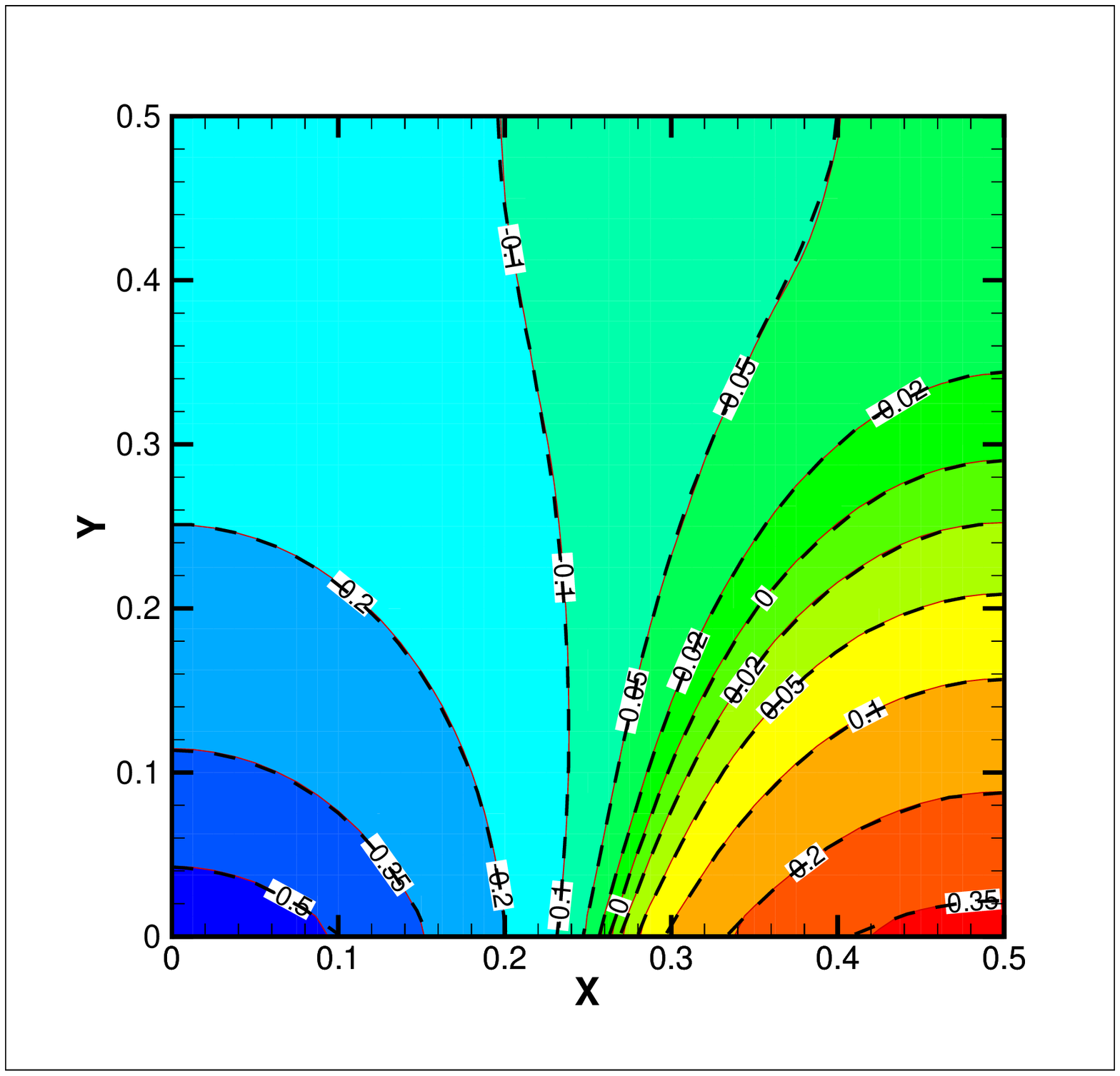}
    }
    \caption{The temperature contour of a spatially boundary temperature variation problem. The dash line is DSMC data extracted
    from the reference \cite{Radtke2011}.}
    \label{fig:VBT} 
\end{figure}
The last simulation is about the response of a gas to a
spatially varying boundary temperature in 2-D domain.
The numerical setup is the same as the case in \cite{Radtke2011}.
Gas is confined between two horizontal boundaries.
The lower boundary at $y=0$ is fully diffusive with a
temperature given by $T_w=T_0(1-0.5\cos(2\pi x))$. An identical
boundary is located at $y=1$. The Knudsen number based
on the separation between the two boundaries is $\mbox{Kn}=1$.
Working gas is argon with reference viscosity defined by Eq.
(\ref{eq:refvis}).
Owing to the symmetries in the x and y directions,
the simulation domain is chosen as $[0, 1/2]\times[0, 1/2]$.
Figure \ref{fig:VBT} shows the temperature contour from ES-model
and S-model. The background data is extracted from reference \cite{Radtke2011}.
Unlike the previous two test cases, ES-model predicts more accurate results
in comparison with DSMC results.

\section{Analysis}

The individual ES-model and S-model were constructed to set the Prandtl number as a free parameter. Physically,
the Prandtl number of a monatomic gas has a fixed value, especially in the continuum flow regime.
Therefore, the Prandtl number should not be taken as a free parameter for monatomic ideal gas.
With a fixed Prandtl number, theoretically there is not any freedom in the ES-model and S-model.
In this study we proposed a generalized kinetic model. Besides a fixed Prandtl number for monatomic gas, the new model
provides one more free parameter.
This free parameter can present a continuum spectrum of kinetic models
with correct Prandtl number.
This parameter provides ways to mimic more complicated physical relaxation process.
With certain choices of this free parameter,  say $C_{es}$,
the S-model and ES-model become a subset of the new model.
In the force driven Poiseuille
flow and unsteady boundary heating problem, the new model provides a way to
get accurate results when $C_{es}$ is set to be larger than 0.

As mentioned in last section, $C_{es}$ and $C_{shak}$ are related to the relaxation of
moments of the distribution function. To shed light on this topic, we exam the Boltzmann
collision term for VHS molecule. The Boltzmann equation is written as following,
\begin{equation}
    \frac{\partial (nf)}{\partial t}+\mathbf{u}\cdot\frac{\partial (nf)}{\partial \mathbf{x}} =
    \mathcal{J}(nf),
    \label{eq:kineticModel}
\end{equation}
where $f$ is normalized distribution function, $n$ represents the particle number density, and
$\mathcal{J}(nf)$ denotes the Boltzmann collision term.
The collision integral is defined as
\begin{equation}
\Delta [Q] = \int_{-\infty}^{+\infty}\int_{-\infty}^{+\infty}\int_{0}^{4 \pi}
  n^2Q(f^*f^*_1-ff_1)c_{r}\sigma d\Omega d\mathbf{u}d\mathbf{u_1},
\label{eq:collisionIntegral}
\end{equation}
where $\Omega$ is the solid angle for scattering molecule, $c_r$ is the relative velocity
between two colliding molecule, and $\sigma$ is the collision cross section.
Consider a spatially homogenous monatomic gas problem.
The moments equation of the Boltzmann equation gives the relaxation process of the moments.
For the quantity $Q$, the relaxation process can be written as
\begin{equation}
\frac{m\partial <nf,Q>}{\partial t} = m\Delta[Q],
\end{equation}
where $<nf,Q> = \int nfQ d\mathbf{u}$, and $m$ is the mass of molecule.
For example, if $Q = u^2$ and for Maxwell molecule, i.e., $\mu \sim T $,
the corresponding relaxation equation is
\begin{equation}
\frac{\partial P_{11}}{\partial t} = \frac{\partial p_{11}}{\partial t} = m\Delta[u^2].
\end{equation}
The collision integral can be obtained explicitly for Maxwell molecule \cite{Bird94}, such as
\begin{equation}
\frac{\partial p_{11}}{\partial t} = \frac{p}{\mu} p_{11}.
\end{equation}
For other molecules, there is no explicit solution. However,
some qualitative results can be deduced from a given distribution function.
Here, we consider two kinds of distribution functions for VHS molecule.
The diameter of VHS molecule is given by
\begin{eqnarray}
d = d_{ref}(c_{r,ref}/c_r)^{\upsilon},
\end{eqnarray}
where $\upsilon = \omega-1/2$.
The first distribution function is the one employed in Grad's thirteen moments method \cite{Harris2004}, and it reads
\begin{eqnarray}
f &=& \mathcal{M}[f]\left(1+(\mathbf{u}-\mathbf{U})\cdot\frac{\mathbf{P}-p\mathbf{I}}{2pRT}\cdot(\mathbf{u}-\mathbf{U})\right.\nonumber \\
& & \left.+\frac{\mathbf{q}}{pRT}\cdot(\mathbf{u}-\mathbf{U})(\frac{\mathbf{(u-U)}^2}{5RT}-1)\right). \label{eq:generalF}
\end{eqnarray}
For the case when ${(\mathbf{P}-p\mathbf{I})}/{(2pRT)}$ and $\mathbf{q}/{(pRT)}$ are much less than $1$,
by substituting Eq.(\ref{eq:generalF}) into collision integral (Eq.(\ref{eq:collisionIntegral})), the above distribution function gives
\begin{eqnarray}
\Delta[\mathbf{uu}]&& \\\label{eq:CIusquare}
&\!\!\!\!\!=& -(n/m)\frac{\sigma_{ref} c_{r,ref}^{2\upsilon}}{2}
\frac{16}{15\sqrt{\pi}}4^{-\upsilon}(RT)^{\frac{1}{2}-\upsilon}\Gamma(4-\upsilon)\mathbf{p},\nonumber \\
\Delta[\frac{1}{2}\mathbf{uu^2}] \\
&\!\!\!\!\!=& -(n/m)\frac{\sigma_{ref}c_{r,ref}^{2\upsilon}}{2}\frac{16}{15\sqrt{\pi}} 4^{-\upsilon}(RT)^{1/2-\upsilon}\Gamma(4-\upsilon) \frac{2}{3}\mathbf{q}, \nonumber
\end{eqnarray}
where $\Gamma$ denotes the Gamma function.
The viscosity of the VHS molecule and the mean collision rate ($1/\tau$)
per molecule in an equilibrium gas of VHS molecules are given by \cite{Bird94}.
Here we reformulate them as following,
\begin{eqnarray}
\mu &=& \frac{15m\sqrt{\pi}4^{\upsilon}(RT)^{1/2+\upsilon}}{8\Gamma(4-\upsilon)\sigma_{ref}c_{r,ref}^{2\upsilon}}, \\
\frac{1}{\tau} &=& 4 n c_{r,ref}^{2\upsilon} \sigma_{ref} 4^{-\upsilon}(RT)^{1/2-\upsilon}\Gamma(2-\upsilon)/\sqrt{\pi}
\end{eqnarray}
Using the above results, the relaxation process of moments of the Boltzamnn equation can be written as,
\begin{eqnarray}
\frac{\partial nf}{\partial t} &=& -\frac{1}{\tau}(nf-(\tau\mathcal{J}(nf)+nf)), \\
\frac{\partial p_{ij}}{\partial t} &=& -\frac{p}{\mu}p_{ij}, \\
\frac{\partial q_{i}}{\partial t} &=& -\frac{2}{3}\frac{p}{\mu}q_{i}.
\end{eqnarray}
Actually, for VHS molecule in a local equilibrium state, the $C_{es}$ can be derived as \cite{Bird94},
\begin{eqnarray}
\frac{1}{\tau}&=&\frac{30}{(7-2\omega)(5-2\omega)}\frac{p}{\mu},\\
C_{es}&=&1-\frac{(7-2\omega)(5-2\omega)}{30}.
\end{eqnarray}
Here $C_{es}$ is confined in a domain of $[0.2, 0.5]$ for VHS molecules, and can be taken as
a constant. However, there is no universal conclusion.
For the shock structure calculation,
the new model with such a range of $C_{es}$ seems to give inappropriate solutions.

Hereafter we consider another distribution function. Assume the distribution function is
composed of two delta function, say,
\begin{eqnarray}
f = \alpha \delta(u-(1-\alpha)u_0) + (1-\alpha) \delta(u + \alpha u_0),\label{eq:pdfDelta}
\end{eqnarray}
where $\alpha\in[0,1]$, $u$ denotes molecule velocity in $x$ direction. Then the pressure, stress
tensor and heat flux can be expressed by $\alpha$ and $u_0$,
\begin{eqnarray}
p &=& \frac{1}{3}mn\alpha(1-\alpha)u_0^2, \\
P_{11} &=& 3p , \\
q_1 &=& \frac{1}{2}mn\alpha(1-\alpha)(1-2\alpha)u_0^3 .
\end{eqnarray}
The collision rate is
\begin{eqnarray}
\frac{1}{\tau} = & 2 n \pi d_{ref}^2 c_{r,ref}^{2\upsilon}\alpha(1-\alpha) u_0^{1-2\upsilon}, &\ 0\leq \upsilon < 1/2, \\
\frac{1}{\tau} = & n \pi d_{ref}^2 c_{r,ref}, &\ \upsilon = 1/2 .
\end{eqnarray}
Note that the ${1}/{\tau}$ is not continuous when $\upsilon = 1/2$, because
two molecules with identical velocity collide with each other with infinite collision cross section between them.
It is inappropriate to count this kind of collision.
So we will discuss the case of $0\leq \upsilon < 1/2$. Substituting Eq.(\ref{eq:pdfDelta}) into Eq.(\ref{eq:collisionIntegral}),
the collision terms give,
\begin{eqnarray}
m\Delta[u^2] &=& -\frac{1}{\tau} p_{11}\frac{(q_1/\rho)^2+(P_{11}/\rho)^3}{(P_{11}/\rho)^3} , \\
m\Delta[\frac{1}{2}u\mathbf{u}^2] &=& -\frac{1}{\tau} \frac{2}{3}q_{1}\frac{(q_1/\rho)^2+(P_{11}/\rho)^3}{(P_{11}/\rho)^3}.
\end{eqnarray}
Here, the relaxation process is totally different from the near equilibrium state as shown before.
$C_{es}$ in this case can be formally written as
$$C_{es} = 1-\frac{(q_1/\rho)^2+(P_{11}/\rho)^3}{(P_{11}/\rho)^3}.$$
Obviously, it is not a constant. Furthermore, it is less than 0 and
can even go to minus infinity.
\begin{figure}[h]
\centering
    \parbox[t]{0.6\textwidth}{
    \includegraphics[totalheight=8cm, bb = 90 35 690 575, clip =
    true]{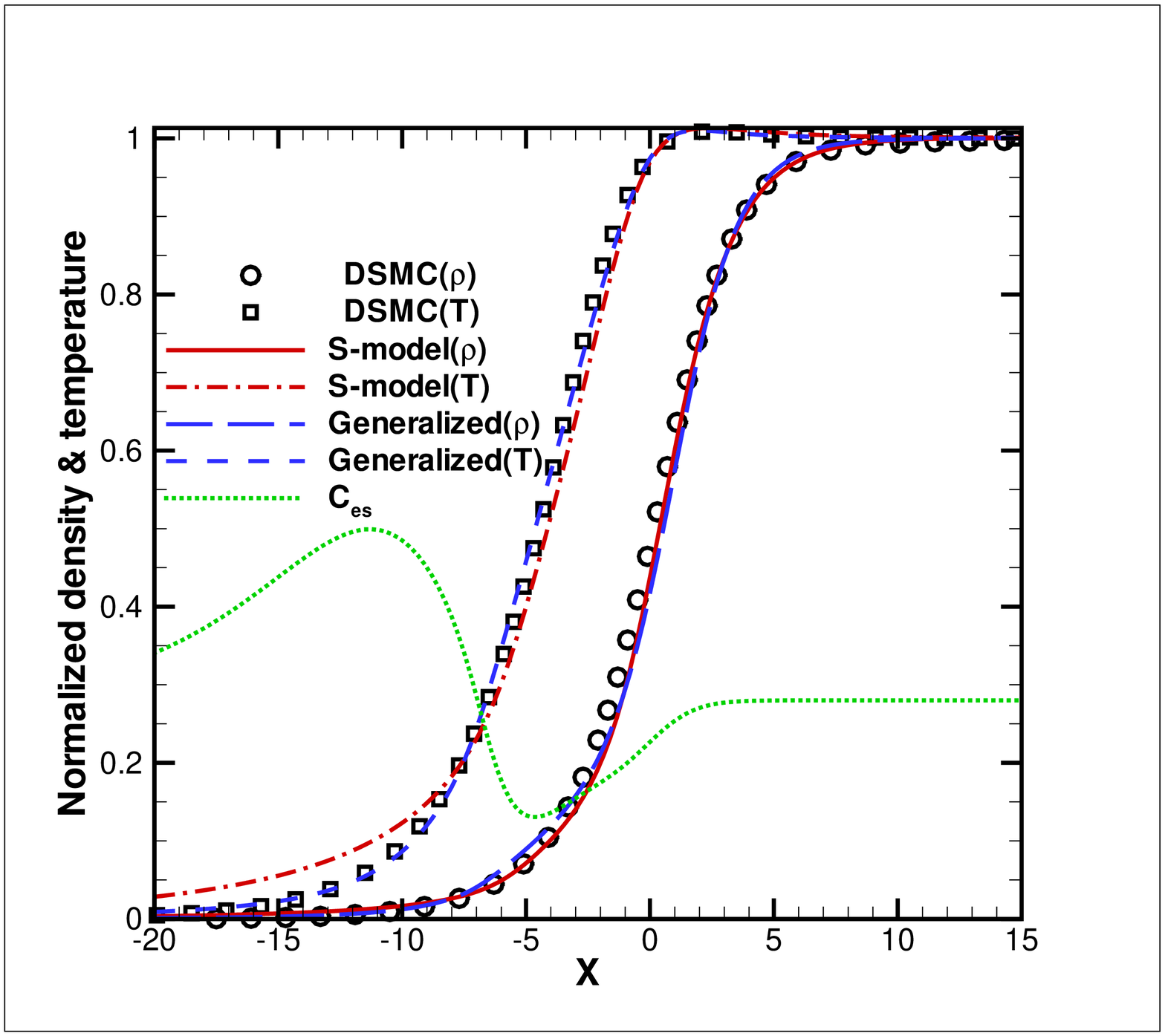}
    }
    \caption{The shock structure from the generalized kinetic  model with a variable $C_{es}$ at $\mbox{Ma} = 8$, and $\omega = 0.81$.}
    \label{fig:shockGood} 
\end{figure}

With this understanding, we construct a variable $C_{es}$ in the shock structure calculation
in order to get a good agreement with DSMC.
As show in figure \ref{fig:shockGood}, a perfect shock structure can be obtained and the
corresponding $C_{es}$ is plotted for this calculation.
The temperature profile is much improved, while the density profile changes only a little bit.
The early raising of temperature in the upstream is suppressed efficiently.

\begin{figure}[h]
\centering
    \parbox[t]{0.6\textwidth}{
    \includegraphics[totalheight=7cm, bb = 90 35 690 575, clip =
    true]{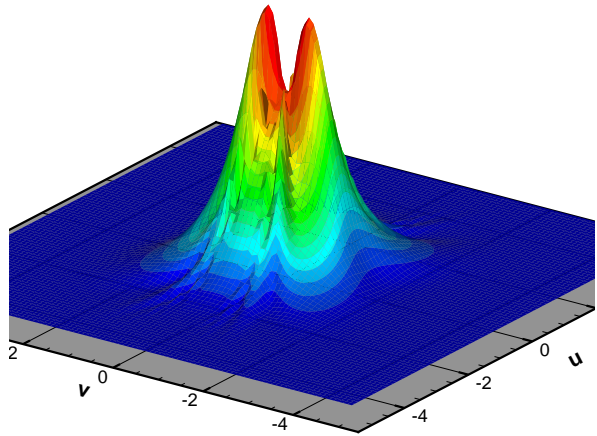}
    }
    \caption{The distribution function at location $(0, 0.5)$ for spatially varying boundary temperature heating problem.}
    \label{fig:VHTpdf} 
\end{figure}

For boundary temperature variation problem, the value of $C_{es}$ is preferred to recover the ES-model,
namely, $C_{es} = -0.5$.
The figure \ref{fig:VHTpdf} shows the distribution function at $(0, 0.5)$.
Based on the above analysis, two peak structure
corresponds to a negative value of $C_{es}$. Therefore, we qualitatively conclude that the ES-model is more appropriate for
this problem. Based on these numerical results and analysis, we believe that this new free parameter has significant
physical insight which deserves its further study.

As mentioned above, the relaxation rate of different moments depends on the distribution function and molecular types.
And $C_{es}$ cannot be taken as a constant for transition flow. For the two-peak distribution functions, this
coefficient could be even far less than 0.
But $C_{es}$ in the ES-model is always constrained in the interval $[-0.5, 1)$ in order to
keep a positive eigenvalue of $\mathbf{T}$.
In fact, an alternative of Gaussian distribution can be adopted in the kinetic model,
\begin{equation}
\mathcal{G}[f] \approx \mathcal{M}[f] (1+(\mathbf{u}-\mathbf{U})\cdot
  \mathbf{T'}\cdot (\mathbf{u}-\mathbf{U})),
\end{equation}
where
\begin{equation}
T'_{ij}=\frac{1}{2(RT)^2}T_{ij}, \quad i\neq j ,
\end{equation}
and
\begin{equation}
T'_{ij}= \frac{1}{2(RT)^2}(T_{ij}-\mbox{trace}(\mathbf{T})/3), \quad i = j .
\end{equation}
Surprisingly, although the above expansion cannot guarantee the positivity of the distribution function,
the numerical results from the above expansion are very close to that where a full Gaussian distribution function is used.
It indicates that for any formulation we adopt in the model equation, the results of macroscopic variables
will be the same, as long as the moments of the collision term are identical from different kinetic models.
Furthermore, replaced by the expansion, the lower bound of $C_{es}$ for Gaussian distribution
can be removed. We can use a value of $C_{es}$ less than $-0.5$.



\section{Conclusion}

In this paper, we have developed a generalized kinetic model through the combination of ES-model and
S-model. With a fixed Prandtl number, this new model provides an additional free parameter, which can be
used to recover the physical solution more accurately. By changing the free parameter the different relaxation time
between different moments of a distribution function can be simulated.
The unified gas kinetic scheme is used for the construction of numerical solution of the generalized kinetic model.
With the variation of this
free parameter, the new model covers the BGK model, ES-model and Shakhov model.
At the same time, it provides a continuum spectrum of kinetic models and different dynamics with a variation of this parameter.
In most cases, the S-model presents more accurate  numerical results.
The numerical study indicates that the essential property for a kinetic model to
capture physically valid solutions is the ratios between the
relaxation rates of different moments of a distribution function.
We believe that the introduction of this generalized kinetic model is important in the study of non-equilibrium flow, and this
free parameter has significant physics basis, which deserves its further study.


\vspace{3ex} {\textbf{Acknowledgments}} \vspace{1ex}

We would like to thank Yonghao Zhang and Lei Wu for their helpful comments
and suggestions. This work was supported by Hong Kong Research Grant Council (621709, 621011),
and grants SRFI11SC05 and RPC10SC11 at HKUST, and
the National Natural Science Funds for Distinguished Young Scholar group under Grant No. 11221061.

\bibliography{reference_bib/ESpShak}

\end{document}